\documentclass[aps,prd,twocolumn,showpacs,eqsecnum,
superscriptaddress]{revtex4-1}

\usepackage[centertags]{amsmath}
\usepackage{amssymb}
\usepackage{latexsym}
\usepackage{enumerate}
\usepackage{graphicx}
\usepackage{mathrsfs}
\usepackage{hyperref}
\usepackage{stmaryrd}
\usepackage{import}
\usepackage{tensor}
\usepackage{color}
\usepackage{bm}
\usepackage{amscd}
\usepackage{tikz}
\usetikzlibrary{snakes}
\usepackage[caption=false]{subfig}
\usepackage{floatrow}
\allowdisplaybreaks[3]
\graphicspath{{./},{figures/}}

\newcommand{\bi}{\begin{itemize}}
\newcommand{\ei}{\end{itemize}}
\newcommand{\be}{\begin{equation}}
\newcommand{\ee}{\end{equation}}

\renewcommand{\a}{\alpha}
\renewcommand{\b}{\beta}
\newcommand{\g}{\gamma}
\newcommand{\G}{\Gamma}
\renewcommand{\d}{\delta}

\renewcommand{\O}{\Omega}

\renewcommand{\th}{\theta}

\begin{document}

\title{Trumpet Initial Data for Boosted Black Holes}

\author{Kyle Slinker}
\author{Charles R. Evans}
\affiliation{Department of Physics and Astronomy, University of North 
Carolina, Chapel Hill, North Carolina 27599, USA}
\author{Mark Hannam}
\affiliation{School of Physics and Astronomy, Cardiff University, Queens 
Building, CF24 3AA Cardiff, United Kingdom}

\begin{abstract}
We describe a procedure for constructing initial data for boosted black holes 
in the moving-punctures approach to numerical relativity that endows the 
initial time slice from the outset with trumpet geometry within the black 
hole interiors.  We then demonstrate the procedure in numerical simulations 
using an evolution code from the \texttt{Einstein Toolkit} that employs 
1+log slicing.  By using boosted Kerr-Schild geometry as an intermediate 
step in the construction, the Lorentz boost of a single black hole can be 
precisely specified and multiple, widely separated black holes can be treated 
approximately by superposition of single hole data.  There is room within 
the scheme for later improvement to resolve (iterate) the constraint 
equations in the multiple black hole case.  The approach is shown to yield 
an initial trumpet slice for one black hole that is close to, and rapidly 
settles to, a stationary trumpet geometry.  By avoiding the assumption of 
conformal flatness, initial data in this new approach is shown to contain 
initial transient (or ``junk'') radiation that is suppressed by as much as 
two orders of magnitude relative to that in comparable Bowen-York initial 
data.

\end{abstract}


\maketitle

%\tableofcontents

\section{Introduction}
\label{sec:intro}

The first Advanced LIGO observations 
\cite{LVC1602.03837,LVC1606.04855,LVC1706.01812,LVC1709.09660,LVC1711.05578} 
of merging black hole 
binaries have ushered in a new era of gravitational wave astronomy.  These 
observations have already revealed a new class of heavy stellar-mass 
black holes \cite{LVC1602.03840}.  Further analysis has provided tests of 
general relativity \cite{LVC1602.03841,YuneYagiPret16,LVC1706.01812}, in part 
constraining the mass of the graviton and other potential gravitational-wave 
dispersion effects.  Detection and subsequent physical parameter estimation 
\cite{LVC1602.03840,LVC1606.04856} 
utilize increasingly sophisticated theoretical waveform 
models~\cite{KhanETC16,Hannam:2013oca,TaraETC14,BoheETC17}. 
These are constructed from a combination of post-Newtonian~\cite{Blan14}, 
effective-one-body~\cite{BuonDamo99} and 
perturbation-theory results, and calibrated to numerical-relativity simulations.
%utilize increasingly sophisticated 
%theoretical waveform templates that are derived from post-Newtonian 
%calculations, effective-one-body models, frequency-domain-based 
%phenomenological models \cite{HusaETC16,KhanETC16}, black hole perturbation 
%theory calibrations, or numerical relativity simulations \cite{LVC1606.01262}.
While computationally intensive, numerical relativity has the advantage of 
providing self-consistent waveforms that follow the black hole binary 
evolution all the way from (late) inspiral through merger and ringdown.  
Several numerical relativity approaches are employed, with efforts to ensure 
that computed waveforms are accurate for matching and parameter estimation 
purposes (see \cite{LVC1611.07531} for comparisons to the GW150914 waveform).

One particular successful numerical relativity formulation for modeling black 
hole encounters is the moving-punctures method 
\cite{CampETC06,BakeETC06a} (see also \cite{BaumShap10}).  In this scheme, 
the coordinate conditions, and specifically the 1+log slicing 
condition, are known to produce at late times spacelike slices that exhibit 
\emph{trumpet} geometry \cite{HannETC07b,BaumNacu07,HannETC08,Brow08,Brug09}.  
These trumpet slices smoothly penetrate the horizon 
and limit in the black hole interior on a two-surface of non-zero 
proper area where the lapse function vanishes.  Thus the surface surrounds 
and maintains separation from the enclosed spacetime singularity.  This 
boundary of the spatial slice is nonetheless separated by an infinite proper 
distance from any point on the rest of the time slice.  The coordinate 
conditions furthermore map this entire limiting two-surface to a single 
(moving) point in the spatial coordinate system, which marks the puncture.  
Viewed in the context of a Penrose diagram, the trumpet slices of 
Schwarzschild spacetime reach from spatial infinity in ``our'' universe to 
future timelike infinity $i^{+}$ on the other side of the wormhole.  

Simulations using the moving-punctures method typically employ Bowen-York 
initial data \cite{BoweYork80,BranBrug97} to satisfy the constraint 
equations.  The Bowen-York procedure simultaneously finds consistent starting 
data and constructs the initial spatial slice, which is a symmetric wormhole 
through the black hole interior whose geometry contrasts sharply with that 
of trumpet slices.  Symmetric wormhole slices cannot be stationary if the 
lapse is to be everywhere positive \cite{HannETC03} and it is now well 
understood \cite{Brow08,HannETC08} how the 
moving-punctures gauge conditions act to evolve the data on the initial 
wormhole slice, drawing spatial grid points toward trumpet slices.  One 
minor byproduct of the inconsistency between initial and late-time slice 
geometry is that the velocity of the puncture initially vanishes even for a 
black hole with nonzero linear momentum.  As the trumpet slices develop and 
settle toward stationarity, the puncture rapidly accelerates to a velocity 
consistent with its linear momentum.  A more serious artefact of using 
Bowen-York data stems from the conformal flatness of its three-geometry.  
This assumption introduces spurious initial transient gravitational 
waves, commonly called junk radiation.  Depending upon the simulation, junk 
radiation can partly overlap the physical waveform, interfering with the 
extraction of the latter.  Conformal flatness and junk radiation also 
severely limit specifiable black hole spin ($\chi \lesssim 0.93$) 
\cite{CookYork90,DainLousTaka02,DainLousZloc08,HannHusaOMur09} 
and affect the ability to specify a black hole's 
boost~\cite{YorkPira82,CookYork90}.
It is possible to produce Bowen-York puncture black holes with a trumpet 
geometry~\cite{HannHusaOMur09}, but the problems of zero initial coordinate
speed and junk radiation remain. 

This paper presents an alternative to Bowen-York initial data for specifying 
boosted black holes in the moving-punctures formalism.  The new approach 
avoids conformal flatness and constructs its data directly on a time slice 
with trumpet geometry.  When used for one black hole in an evolution code 
with 1+log slicing and the $\Gamma$-driver coordinate conditions, the evolution 
settles rapidly to a near-stationary black hole with momentum consistent with 
the initially specified Lorentz factor.  The elimination of assumed conformal 
flatness brings with it decreases in junk radiation by as much as two orders 
of magnitude.  The present approach bears some similarity to that of 
\cite{LoveETC08} who used superposition of conformally Kerr black holes in 
the Kerr-Schild gauge for simulations with the SXS Collaboration 
\cite{Spec} code, and were able to evolve black holes with spins as high as 
$0.994$ \cite{LoveETC12,ScheETC15}.  Our present effort is directed to 
improving the construction of boosted (but non-spinning) black holes, which 
might allow further exploration of high energy encounters between grazing 
holes \cite{Sper07,SperETC08,SperETC09,HealETC16}.  

Prior work \cite{HannETC07a} on highly spinning black holes showed the 
importance of eliminating the conformal flatness assumption in reducing junk 
radiation.  In contrast, merely adding a trumpet slice modification to 
Bowen-York data \cite{HannHusaOMur09}, which preserved conformal flatness, 
had almost no effect on the junk radiation.  We conclude that the sharp 
reduction we see in junk radiation is due to adopting boosted Kerr-Schild 
coordinates as an intermediate step in the construction, with the additional 
transformation to a trumpet slice serving to avoid encountering the future 
singularity.  

Other recent works on boosted or spinning black holes include \cite{RuchETC14,
Ruch15,HealETC16}, where similar improvements are found in junk radiation by 
eliminating conformal flatness with approaches that are different from ours in 
detail.  Kerr-Schild coordinates were first employed by \cite{MatzHuqShoe98} to 
construct boosted Schwarzschild holes.  Analytic solutions for static black 
holes with trumpet slices were presented in \cite{DennBaum14,DennBaumMont14}.  
Our approach extends the idea of constructing an initial trumpet slice, but 
does so in a specific way, incorporating a boost through application of an 
intermediate Kerr-Schild coordinate system and drawing upon \cite{HannETC08} 
for a height function to build-in the trumpet.

The outline of this paper is as follows.  In Section \ref{sec:bt_coordinates}  
we describe our new means of constructing boosted-trumpet initial data, 
including a review of the unboosted case given previously in \cite{HannETC08} 
that factors into our more general treatment.  In 
Sec.~\ref{sec:numerical_setup}, we discuss the numerical implementation of the 
scheme in a moving-punctures code from the \texttt{Einstein Toolkit}, the 
evolutionary gauge conditions used, and the preparation of Bowen-York initial 
data for a single black hole that is used as a control.  We then utilize the 
new initial data in Sec.~\ref{sec:results} in simulations to make side-by-side 
comparisons with runs using comparably specified (boosted) Bowen-York data.  
In that section we also describe a set of diagnostic tools (e.g., asymptotic 
mass and momentum, horizon measures, and Newman-Penrose $\psi_4$ assessment of 
emitted gravitational radiation) and adjustments needed to analyze a single 
moving black hole.  Our conclusions are drawn in Sec.~\ref{sec:conclusions}.  
Throughout this paper we set $c = G = 1$, use metric signature 
$\left(-+++\right)$ and sign conventions of Misner, Thorne, and Wheeler 
\cite{MisnThorWhee73}, and largely adhere to standard numerical relativity 
\cite{BaumShap10} notation.

\section{Trumpet Coordinates for Static and Boosted Black Holes}
\label{sec:bt_coordinates}

A natural way to introduce a boost to a black hole is to use Kerr-Schild 
(KS) coordinates in which the line element takes the form
\begin{equation}
ds^2 = {g'}_{ab} \, {dx'}^a {dx'}^b  
= \left( \eta_{ab} + 2 H \, {k'}_a {k'}_b \right) {dx'}^a {dx'}^b , 
\label{eqn:SchKSlineelement}
\end{equation}
where ${k'}_a$ is (ingoing) null in both the background metric $\eta_{ab}$ and 
full metric ${g'}_{ab}$.  Prime indicates the static (meaning here unboosted) 
frame, in which a Schwarzschild black hole has $H = M/R$, with $R$ being the 
areal radial coordinate, and ${k'}_a = \left(1,x'/R,y'/R,z'/R\right)$.  
Changing from rectangular to spherical-polar form, this metric takes on the 
familiar ingoing Eddington-Finkelstein (IEF) form.  Left in rectangular form, a 
boost can be introduced -- despite spacetime curvature -- by merely making a 
global Lorentz transformation 
\begin{equation}
x^a = {\Lambda^a}_b \, {x'}^b , \qquad 
k_a = {\Lambda_a}^b \, {k'}_b , 
\end{equation}
which allows the boosted metric to remain form-invariant,
\begin{equation}
\label{eqn:KSforminvariant}
g_{ab} = \eta_{ab} + \frac{2M}{R} {k}_a {k}_b .
\end{equation}
See \cite{MatzHuqShoe98} for early discussion of KS coordinates in the 
context of numerical relativity.

Unfortunately, both the unboosted (IEF) and boosted Kerr-Schild coordinates 
have time slices that intersect the singularity in the black hole interior. 
So unless the interior is excised (as the SXS Collaboration does), 
Kerr-Schild coordinate systems by themselves are unsuited for specifying 
initial data in moving-punctures calculations.  They do, however, serve in 
this paper as intermediate coordinate systems that allow us to introduce a 
boost, giving the black hole a precisely controllable momentum.

Our approach to generating boosted trumpet slices is relatively easy to state. 
We begin with a Schwarzschild black hole in IEF (rectangular) coordinates 
(KS with $v=0$) and then boost to KS form with a specified Lorentz factor.  
We next transform from KS coordinates by subtracting from the KS time 
coordinate $\overline{t}$ a suitable (to be defined) height function to 
introduce a spacelike trumpet surface.  Finally, we transform spatial 
coordinates to quasi-isotropic form to map the moving trumpet limit surface 
to a moving point (i.e., the puncture).  The height function we use is 
constructed following \cite{HannETC08}, and is an exact match for the 
simulation gauge conditions (1+log slicing and the $\Gamma$-driver condition)
only in the case of a static black hole.  When used with a moving black hole, 
the initial trumpet slice is an approximation to the surface that emerges at 
later times in simulations.  Despite the approximation, in simulations we 
find that the coordinates settle after a few light crossing times and 
relatively high single black hole speeds ($v \lesssim 0.85$) can be 
successfully specified and 
evolved.  In the balance of this section, we describe our procedure for 
generating boosted trumpet data (in Sec.~\ref{subsec:procedure}) but first 
review (in Sec.~\ref{subsec:HHOBO_results}) the construction of the static 
black hole trumpet, following Ref.~\cite{HannETC08}.

\subsection{Review of Trumpet Slicing a Static Black Hole}
\label{subsec:HHOBO_results}

Here we consider the transformation from IEF to trumpet coordinates for a 
static black hole.  Later we will consider several other coordinate systems 
in order to outline our full procedure.  Accordingly, it is important to 
set out a notation to distinguish these coordinate systems.  In what follows, 
we use a prime to refer to any static coordinate system.  In parallel, we 
will use a bar to refer to any KS coordinate system, either boosted or 
not (IEF).  Coordinates lacking a bar will denote those with trumpet slices 
and with isotropic or quasi-isotropic mapping of the trumpet end to a 
puncture.  Thus, IEF coordinates in rectangular form are denoted by 
$(\overline{t}^\prime,\overline{x}^{\prime i})$, while boosted KS coordinates 
are written as $(\overline{t},\overline{x}^i)$ (no prime).  Our final 
coordinates for the boosted trumpet data will be simply $(t,x^i)$.  The 
transformation reviewed in this section takes a static black hole in IEF to 
static trumpet coordinates, with the latter coordinates designated by 
$(t',x^{\prime i})$ (or its spherical-polar variant).

Our brief summary diverges from \cite{HannETC08} slightly, by beginning with 
IEF coordinates 
\begin{align}
ds^2&=-fd\overline{t}^{\prime2}+\frac{4M}{R}
d\overline{t}^\prime dR
\nonumber\\
{}&\qquad\qquad{}+\left(1+\frac{2M}{R}\right)dR^2+R^2d\O^2 ,
\label{eqn:InEFlineelement}
\end{align}
where $f\equiv1-2M/R$.  A new time coordinate $t^\prime$ is introduced 
\begin{equation}
t^\prime=\overline{t}^\prime-h_s(R) ,
\label{eqn:HHOBO_height}
\end{equation}
where $h_s(R)$ is an as yet undetermined spherically-symmetric height 
function for the static case.  Following this change the line element 
(\ref{eqn:InEFlineelement}) becomes
\begin{align}
ds^2&=-fdt^{\prime2}-2\left[f\frac{dh_s}{dR}-\frac{2M}{R}\right]dt^\prime 
dR
\nonumber
\\
{}&\qquad{}+\left[1+\frac{2M}{R}+\frac{4M}{R}\frac{dh_s}{dR}
-f\left(\frac{dh_s}{dR}\right)^2\right]dR^2\nonumber\\
{}&\qquad{}+R^2d\O^2 .
\label{eqn:firstHHOBOtransformation}
\end{align}
Given time-translation symmetry, the metric depends on $dh_s/dR$ but not 
on $h_s$ itself.  In stationary spacetimes, introducing a transformation in 
the time coordinate with a time-independent height function is a standard 
technique.  Hannam \emph{et al}.~\cite{HannETC08} were the first to find 
the height 
function to match 1+log slicing.  Their height function only differs from 
ours because they began with Schwarzschild coordinates while we began with 
IEF coordinates.  The lapse function, shift vector, three-metric, slice 
normal $n_a$, and extrinsic curvature can be easily read off or determined 
from \eqref{eqn:firstHHOBOtransformation}.

\begin{figure}
\includegraphics{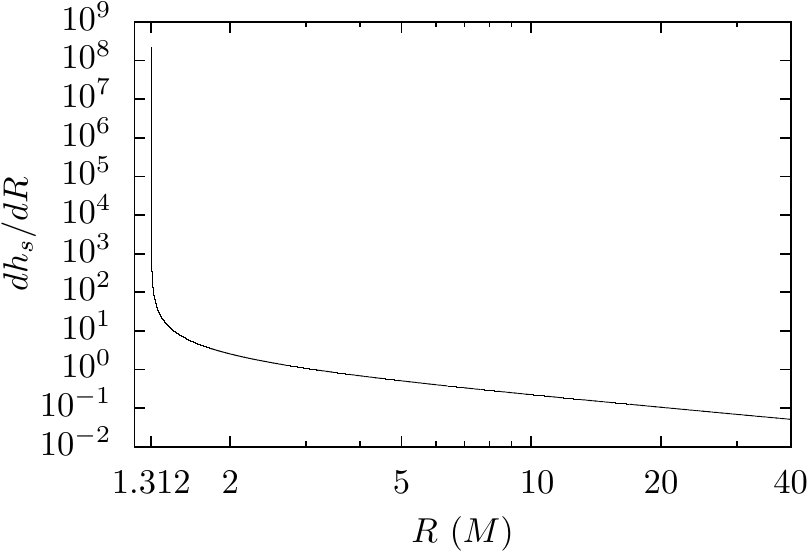}
\caption{Radial derivative of static height function.  The function 
$dh_s/dR$ is shown plotted versus areal radial coordinate $R$ on logarithmic 
scales.  The function grows without bound as the limiting surface 
$R_0 \simeq 1.3124 M$ is approached.  No divergence occurs at the event 
horizon $R = 2 M$ in our case (contrast with \cite{HannETC08}) because we 
transform from IEF coordinates, not Schwarzschild coordinates.}
\label{fig:logdhdR}
\end{figure}

The class of 1+log slicing conditions (for various constants $n$) follow
\begin{equation}
\label{eqn:onepluslogn}
\left(\tensor{\partial}{_t}-\tensor{\b}{^i}\tensor{\partial}
{_i}\right)\a=-n\a K .
\end{equation}
Assuming stationarity then gives a differential equation satisfied by the 
lapse
\begin{equation}
\frac{d\a}{dR}=-\frac{n(3M-2R+2R\a^2)}{R(R-2M+nR\a-R\a^2)} .
\label{eqn:alphaODE}
\end{equation}
The relevant solution is the one that passes through the critical point of 
(\ref{eqn:alphaODE}).  An implicit solution is found to be 
\begin{equation}
\a^2 = 1-\frac{2M}{R}+\frac{C(n)^2e^{2\a/n}}{R^4} ,
\label{eqn:lapsesolution}
\end{equation}
where the constant $C(n)$ is given by
\begin{equation}
C(n)^2 = \frac{[3n+\sqrt{4+9n^2}]^3}{128n^3}e^{-2\a_c/n}M^4 ,
\end{equation}
and where
\begin{subequations}\begin{align}
\a_c^2&=\frac{\sqrt{4+9n^2}-3n}{\sqrt{4+9n^2}+3n},\\
R_c&=\frac{3n+\sqrt{4+9n^2}}{4n}M ,
\end{align}\end{subequations}
give the location of the critical point along the $\a(R)$ curve where the 
numerator and denominator of the right hand side of (\ref{eqn:alphaODE}) 
simultaneously vanish.  In what follows we specialize to the standard 1+log 
gauge condition and set $n=2$, after which the specific numerical values 
$\a_c \simeq 0.16228$, $R_c \simeq 1.5406 M$, and $C(2) \simeq 1.2467 M^2$ are 
found.  Once the lapse function is determined by solving \eqref{eqn:alphaODE}, 
the throat (where $\a = 0$) is located at $R_0 \simeq 1.3124 M$ and 
(\ref{eqn:firstHHOBOtransformation}) can be used to find the required 
derivative of the height function
\begin{equation}\frac{dh_s}{dR}=\frac{2M\a-R\sqrt{\a^2-f}}{\a fR} .
\label{eqn:height_function}
\end{equation}
Fig.~\ref{fig:logdhdR} shows the behavior of $dh_s/dR$ as a function of $R$.

The calculation above reestablishes the known stationary 1+log solution of 
\cite{HannETC08}.  A second (spatial) coordinate transformation is then 
applied to map areal radial coordinate $R$ to an isotropic radial coordinate 
$r^\prime$, and in the process draw the limiting surface of the trumpet to a 
single point.  Assuming $R=R_s(r^\prime)$, \eqref{eqn:firstHHOBOtransformation} 
is transformed while requiring the new three-metric to be isotropic, yielding 
the differential equation
\begin{equation}
\frac{dR_s}{dr^\prime}=\frac{\a R_s}{r^\prime} .
\label{eqn:RODE}
\end{equation}

\subsubsection{Numerical Solution}
\label{subsubsec:integral_ODE}

We seek to solve the coupled system of differential equations 
(\ref{eqn:alphaODE}) and (\ref{eqn:RODE}) as functions of $r^\prime$.  
The required boundary condition on $r^\prime$ in (\ref{eqn:RODE}) is 
$r^\prime /R \rightarrow 1$ as $r^\prime\rightarrow\infty$ and 
$\a\rightarrow 1$.  Unfortunately, the numerical initial-value integration of 
(\ref{eqn:alphaODE}) needs to begin at the critical point in order to ensure a 
smooth passage of the solution there, and at $R_c$ we have no \emph{a priori} 
knowledge of the appropriate starting value of $r'$ to match its 
boundary condition at infinity.  Fortunately, the system of equations is 
autonomous in $\ln r'$, so that for any solution $R_s(r')$ we can scale $r'$ 
arbitrarily and still have a solution.  Following \cite{HannETC08} we express
(\ref{eqn:RODE}) as an integral, integrate by parts, and obtain
\begin{equation}
r^\prime=R^{1/\alpha}\exp\left[\int^\alpha_{\alpha_c}\frac{\ln 
R(\tilde{\alpha})}{\tilde{\alpha}^2}d\tilde{\alpha}-C_0\right] ,
\label{eqn:rprime_sln}
\end{equation}
where the integration is begun at the critical point and the integration 
constant $C_0$ accommodates the arbitrary scaling of $r^\prime$.  The behavior 
$r^\prime /R \rightarrow 1$ then requires
\begin{equation}
\label{eqn:C0integral}
C_0=\int^1_{\alpha_c}\frac{\ln 
R(\tilde{\alpha})}{\tilde{\alpha}^2}d\tilde{\alpha} .
\end{equation}

If we set $C_0 = 0$ in \eqref{eqn:rprime_sln}, the result is an implicit 
solution for a (scaled) isotropic coordinate $\tilde{r} = \tilde{r}(\alpha)$, 
whose critical point value will be $\tilde{r}_c = R_c^{1/\alpha_c}$.  If 
$C_0$ is separately determined, then $r^\prime$ is found via 
$r^\prime = \tilde{r} \exp(-C_0)$.  To find $C_0$ we use \eqref{eqn:alphaODE} 
and \eqref{eqn:RODE} to convert \eqref{eqn:C0integral} into a differential 
equation for a variable $k$ 
\begin{equation}
\frac{dk}{d\tilde{r}} \equiv 
\frac{\ln R}{\alpha^2}\frac{d\alpha}{dR}
\frac{dR}{d\tilde{r}} .
\end{equation}
Following outward integration, 
$C_0 = \lim_ {\tilde{r} \rightarrow\infty} k(\tilde{r})$ if we set 
$k(\tilde{r}_c) = 0$.

In summary, the full system we integrate is
\begin{subequations}
\label{eqn:coupledODEs}
\begin{align}\frac{dR_s}{d\tilde{r}}
&=\frac{\a R_s}{\tilde{r}}\\
\frac{d\a}{d\tilde{r}}&=-\frac{\a R_s}{\tilde{r}}\frac{2(3M-2R_s+2R_s\a^2)}
{R_s(R_s-2M+2R_s\a-R_s\a^2)}\\
\frac{dk}{d\tilde{r}}&=-\frac{\ln R_s}{\a^2}\frac{\a R_s}{\tilde{r}}
\frac{2(3M-2R_s+2R_s\a^2)}{R_s(R_s-2M+2R_s\a-R_s\a^2)} .
\end{align}
\end{subequations}
We integrate from the critical point, both outward to large $\tilde{r}$ and 
(with the first two equations only) inward to $\tilde{r} = 0$.  Following 
integration, we convert from $\tilde{r}$ to $r^\prime$.  To obtain 
the critical solution we use L'H\^opital's rule on (\ref{eqn:alphaODE}) to 
find the derivative $\left.\frac{d\a}{dR}\right|_{R_c}$ 
\begin{align}
{}&\left.\frac{d\a}{dR}\right|_{R_c}\left[2(1-\a_c)R_c^2\right]
=-R_c+M-6R_c\a_c
\nonumber\\
{}&\qquad\qquad{}+R_c\a_c^2+\Big[
8R_c^2(\a_c^2-1)(\a_c-1)\Big.
\nonumber\\
{}&\qquad\qquad\qquad\Big.{}+[M+R_c(\a_c^2-6\a_c-1)]^2\Big]^{1/2} .
\label{eqn:criticalaprime}
\end{align}

Equations (\ref{eqn:coupledODEs}) are solved numerically in Mathematica in 
order to ensure machine double precision when input to our C code.  Lookup 
tables are created for $R_s(r^\prime)$ and its first three derivatives and 
for $dh_s/dR$ and its first two derivatives on a grid of $r^\prime$ values.  
Entries are spaced evenly in $\ln r^\prime$ to provide higher resolution near 
the puncture.  

\subsubsection{Series Expansions}
\label{subsubsec:series_ODE}

It is useful to have series expansions of the static trumpet functions 
$dh_s/dR$, $R_s(r^\prime)$, and $\alpha(R)$ to later examine the asymptotic 
properties of our boosted-trumpet spacetime.  Starting with the $n=2$ 
case of \eqref{eqn:alphaODE} we find the asymptotic expansion
\begin{equation}
\a(R)=1-\frac{M}{R}-\frac{M^2}{2R^2}-\frac{M^3}{2R^3}+
\frac{N-160}{256}\frac{M^4}{R^4}+
\cdots ,
\label{eqn:a(R)_expansion}
\end{equation}
where
\begin{equation}
N\equiv\left(3+\sqrt{10}\right)^3e^{4-\sqrt{10}}
\simeq 540.81 .
\end{equation}
Expansion of (\ref{eqn:RODE}) yields
\begin{equation}
\frac{R_s}{M}=\frac{r^\prime}{M}+1+\frac{M}{4r^\prime}-
\frac{N}{1024}\frac{M^3}{{r^\prime}^3}+\frac{3N}{1280}\frac{M^4}{{r^\prime}^4}
+\cdots .
\label{eqn:R(r)_expansion}
\end{equation}
Together, (\ref{eqn:a(R)_expansion}) and (\ref{eqn:R(r)_expansion}) give
\begin{equation}
\a(r^\prime)=1-\frac{M}{r^\prime}+
\frac{M^2}{2{r^\prime}^2}-\frac{M^3}{4{r^\prime}^3}+
\frac{N+32}{256}\frac{M^4}{{r^\prime}^4}
+\cdots .
\label{eqn:a(r)_expansion}
\end{equation}
The static trumpet lapse function $\a(r^\prime)$ of this section is an 
auxiliary variable useful for computing the trumpet functions $dh_s/dR$ and 
$R_s(r^\prime)$.  In subsequent sections, the boosted trumpet lapse $\a$ will 
be a distinct multi-dimensional function derived by the new procedure.  
Finally, from (\ref{eqn:height_function}) we obtain an expansion of the 
height function derivative,
\begin{equation}
\frac{dh_s}{dR}(r^\prime)=\frac{2M}{r^\prime}-
\frac{\sqrt{2N}-32}{16}\frac{M^2}{{r^\prime}^2}+\cdots.
%\mathcal{O}\left(\frac{M^3}{{r^\prime}^3}\right)
\end{equation}

%The retained terms in the asymptotic expansions of $\a$ and $R_s$ provide 
We retain terms to order $1/{r^{\prime}}^{10}$ so that our expansions of $\a$ 
and $R_s$ show 
agreement with the numerical integrations to better than a part in $10^{13}$ 
for all $r^\prime>50M$.  The expansions are used to predict the asymptotic 
properties of Weyl scalars and Arnowitt, Deser, and Misner (ADM) measures
\cite{ArnoDeseMisn08}, which are compared to the 
boosted trumpet simulations at large radii (e.g., $r^\prime\simeq 100M$). 

\subsection{Trumpet Slicing a Boosted Black Hole}
\label{subsec:procedure}

A boost along the $z$ direction maps the spacetime in (rectangular) IEF 
coordinates $(\overline{t}^\prime,\overline{x}^{\prime i})$ to its form in 
boosted KS coordinates $(\overline{t},\overline{x}^i)$ using a (globally 
applied) Lorentz transformation
\begin{equation}
\overline{t} = \g(\overline{t}^\prime+v\overline{z}^\prime), \quad
\overline{z} = \g(\overline{z}^\prime+v\overline{t}^\prime) ,\quad
\overline{x} = \overline{x}^\prime ,\quad
\overline{y} = \overline{y}^\prime .
\label{eqn:transformation1}
\end{equation}
Following the boost, the line element retains its form 
\eqref{eqn:KSforminvariant} but with transformed null vector
\begin{equation}
\label{eqn:boostednull}
\overline{k}_a = 
\left(\gamma-\frac{v\gamma^2}{R}\left(\overline{z}-v\overline{t}\right),
\frac{\overline{x}}{R},\frac{\overline{y}}{R},-v\gamma+\frac{\gamma^2}{R}
\left(\overline{z}-v\overline{t}\right)\right) ,
\end{equation}
where surfaces of constant $R$ (and $\overline{t}$) are now ellipsoids
\begin{equation}
\label{eqn:ellipsoids}
R^2=\overline{x}^2+\overline{y}^2+\gamma^2\left(\overline{z}-v\overline{t}
\right)^2 .
\end{equation}

With the black hole in a frame in which it appears boosted, we can then seek 
to apply a coordinate transformation analogous to (\ref{eqn:HHOBO_height}) 
between KS time $\overline{t}$ and trumpet coordinate time $t$.  In principle 
one might try to determine the exact height function $h$ that is consistent 
with the 1+log slicing condition \eqref{eqn:onepluslogn} and stationarity.  
Unfortunately, in the present application such a height function would be 
axisymmetric and its determination would require solution of a partial 
differential equation.  Our approach instead is to use the static height 
function $h_s$ (solution of \eqref{eqn:height_function}) as an approximation 
for the initial spacelike surface.  With the initial data surface having at 
least the correct topology, we expected that the coordinates would settle 
rapidly to the late-time stationary behavior in a few light crossing times 
(an assumption since borne out in simulations).  The difference between our 
approach and calculating an exact height function is only a coordinate 
transformation.  We therefore take 
\begin{equation}
\label{eqn:transformation2}
t=\overline{t}-h_s\left(R(\overline{t},\overline{x}^i)\right) ,
\end{equation}
making use of $h_s$ from \ref{subsec:HHOBO_results} but assuming it to be a 
function of the level surfaces of $R$ from \eqref{eqn:ellipsoids}.

\begin{figure}
\includegraphics{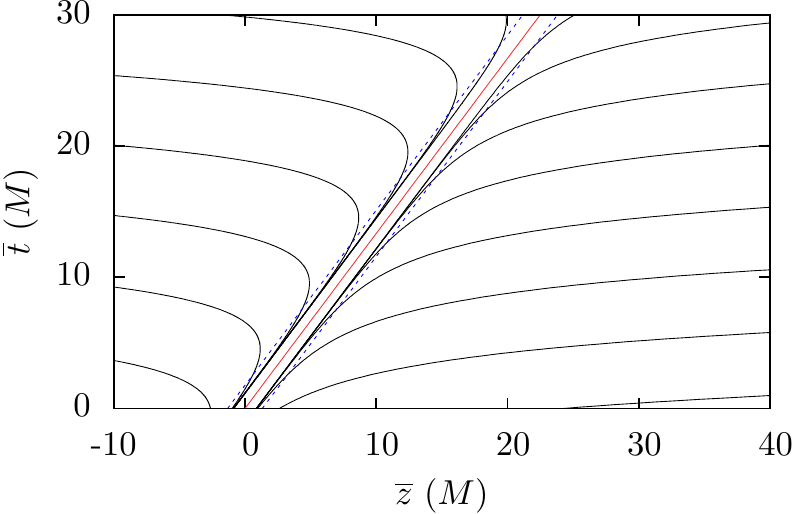}
\caption{Cross section of nested level-surfaces of trumpet time $t$ in the
$\overline{x} = \overline{y} = 0$ plane. The horizontal axis is Kerr-Schild 
coordinate $\overline{z}$ and the vertical axis is KS time coordinate 
$\overline{t}$.  The trumpet surfaces sweep back in $\overline{t}$ time 
within the horizon (blue dashed lines), avoiding the singularity (red line) 
while following its motion.}
\label{fig:foliation_plot}
\end{figure}

The effect of subtracting the height function from KS time as done in 
\eqref{eqn:transformation2} may be visualized by restricting attention to the 
$\overline{x} = \overline{y} = 0$ plane.  In that case, \eqref{eqn:ellipsoids} 
reduces to $R = \gamma \vert \overline{z} - v \, \overline{t} \vert$ and 
\eqref{eqn:transformation2} can be inverted to yield two curves
\begin{subequations}
\begin{align}
\overline{z} &= v \, \overline{t} + \frac{1}{\gamma} \, h_s^{-1} 
( \overline{t} - t) , \qquad {\rm for } \qquad 
\overline{z} > v \, \overline{t} ,
\\
\overline{z} &= v \, \overline{t} - \frac{1}{\gamma} \, h_s^{-1} 
( \overline{t} - t) , \qquad {\rm for } \qquad 
\overline{z} < v \, \overline{t}.
\end{align}
\end{subequations}
Fig.~\ref{fig:foliation_plot} shows a set of the resulting 
nested level-surfaces of trumpet time $t$ plotted in the subspace spanned 
by KS coordinates $(\overline{t},\overline{z})$.  The $t=$ constant surfaces 
form a foliation that surrounds, but avoids, the moving singularity while 
still penetrating the horizon.

For a fixed time $\overline{t}$, \eqref{eqn:ellipsoids} yields a set of 
nested ellipsoids centered on $\overline{x}=0$, $\overline{y}=0$, and 
$\overline{z}=v\overline{t}$.  However, once the trumpet time is introduced 
in \eqref{eqn:transformation2}, for fixed $t$ the ellipsoids diminish in size 
$R$ as $\overline{t}$ becomes increasingly negative, reaching a limit 
$R \rightarrow R_0$.  Effectively the region with KS coordinates such that 
$R < R_0$ is excised from the domain, just as a similar spherical region 
was excised in the static case.  The next step then, also just as in the 
static case, is to map the (moving) limit surface to a single (moving) point 
(i.e., puncture) in a way that is consistent with the transformation 
\eqref{eqn:RODE} from Schwarzschild to isotropic coordinates in spherical 
symmetry.

To make this coordinate transformation, we draw upon the function 
$R_s$ from the static case that follows from solving the differential 
equations \eqref{eqn:coupledODEs} with previously discussed boundary 
conditions.  Here we use $\rho$ as the argument of $R_s(\rho)$, which we 
recall has the properties that $\rho \rightarrow 0$ as $R_s \rightarrow R_0$ 
and $\rho/R_s(\rho) \rightarrow 1$ as $\rho \rightarrow \infty$.  In the 
static case, $\rho/R_s(\rho)$ is the ratio between the new isotropic radial 
coordinate (there called $r^\prime$) and the original Schwarzschild or IEF 
coordinate $R$.  We use this function to define a spatial coordinate 
transformation
\begin{align}
x&=\overline{x} \frac{\rho}{R_s(\rho)} ,
\nonumber\\
y&=\overline{y} \frac{\rho}{R_s(\rho)} ,
\label{eqn:transformation3}
\\
z - v t &= \left(\overline{z} - v \overline{t}\right) \frac{\rho}{R_s(\rho)} ,
\nonumber
\end{align}
taking $(\overline{x},\overline{y},\overline{z})\rightarrow (x,y,z)$ without 
making any additional change to the time coordinate $t$ (here $\overline{t}$ 
is a function of $t$ via (\ref{eqn:transformation2})).  If each equation in 
\eqref{eqn:transformation3} is squared and then combined with 
\eqref{eqn:ellipsoids}, we find that $\rho$ is related to these new spatial 
coordinates by
\begin{equation}
\label{eqn:rhosquared}
\rho^2 = x^2 + y^2 + \g^2 (z - vt)^2 .
\end{equation}
We see that $R=$ constant ellipsoids in KS spatial coordinates, which 
are limited by $R > R_0$ on trumpet slices, map to self-similar $\rho=$ 
constant ellipsoids in the final spatial coordinates, which are centered 
on $x=0$, $y=0$, and $z=v t$ and which squeeze the end of the trumpet to 
that single moving point.

\begin{table}[h!tbp]\begin{center}
\caption{Sequence of transformations for boosted trumpet coordinates. 
The relationship between each system is given to the right of the
 $\downarrow$. 
}
\begin{tabular}{cl}
\hline\hline
{}&Sequence of Coordinate Changes\\
\hline
\smallskip
$\{\overline{t}^\prime,\overline{x}^\prime,\overline{y}^\prime,
\overline{z}^\prime\}$&Ingoing Eddington-Finkelstein (rectangular) 
\\
\smallskip
$\downarrow$&$\overline{t}=\g(\overline{t}^\prime+v\overline{z})
\qquad\overline{z}=\g(\overline{z}^\prime+v\overline{t})$
\\
\smallskip
$\{\overline{t},\overline{x},\overline{y},\overline{z}\}$&
Kerr-Schild (boosted)
\\
\smallskip
$\downarrow$&$t=\overline{t}-h_s(R(\overline{t},\overline{x}^i))$
\\
\smallskip
$\{t,\overline{x},\overline{y},\overline{z}\}$&
(spatial) Kerr-Schild with trumpet slicing
\\
\smallskip
$\downarrow$&$(\overline{x},\overline{y},\overline{z}-v\overline{t}) 
\frac{\rho}{R_s(\rho)} = (x,y,z-v t)$
\\
\smallskip
$\{t,x,y,z\}$&Boosted trumpet slice with moving puncture 
\\
\hline\end{tabular}
\label{table:coordinatechanges}
\end{center}\end{table}

\emph{Summary:}  Constructing a boosted-trumpet coordinate system for a 
single black hole involved three consecutive coordinate transformations.  
The sequence of steps, summarized in Table \ref{table:coordinatechanges}, 
is (1) a Lorentz boost from IEF to KS, (2) addition of a height function to 
transform time and introduce the trumpet, and (3) a map to take the trumpet 
limit surface to the moving puncture.  

Combining (\ref{eqn:transformation1}), (\ref{eqn:transformation2}), and 
(\ref{eqn:transformation3}), we can write the net (reverse) transformation 
from the final boosted trumpet/puncture coordinates to the initial 
rectangular IEF coordinates 
\begin{subequations}
\label{eqn:coordinate_change}
\begin{align}
\overline{t}^\prime&=\g^{-1}[t+h_s(R_s(\rho))]-\g v(z-vt)R_s(\rho)/\rho , \\
\overline{x}^\prime&=xR_s(\rho)/\rho , \\
\overline{y}^\prime&=yR_s(\rho)/\rho , \\
\overline{z}^\prime&=\g(z-vt)R_s(\rho)/\rho .
\end{align}
\end{subequations}

In turn, we can write down (in several parts) the line element in the
new coordinates
\begin{subequations}
\label{eqn:full_line_element}
\begin{align}
ds^2&=-f{d\overline{t}^\prime}^2+\left[dx^2+dy^2+\g^2(dz-vdt)^2
\right]\frac{R_s^2}{\rho^2} \nonumber 
\\
{}&\quad{}+\left(\frac{dR_s}{d\rho}-\frac{R_s}{\rho}\right)
\left(\frac{dR_s}{d\rho}+\frac{R_s}{\rho}\right)
\left(\tensor{\rho}{_,_a}\tensor{dx}{^a}\right)^2 
\\
{}&\quad{}+\frac{4M}{R_s}\frac{dR_s}{d\rho}\left(\tensor{\rho}{_,_a}
\tensor{dx}{^a}\right)\left[d\overline{t}^\prime+\frac{1}{2}\frac{dR_s}{d\rho}
\left(\tensor{\rho}{_,_b}\tensor{dx}{^b}\right)\right] , \nonumber
\end{align}
where the Kerr-Schild coordinate differential $d\overline{t}^\prime$ above 
must be replaced with
\begin{align}
d\overline{t}^\prime&=\frac{dt}{\g}+ 
\frac{1}{\g}\frac{dh_s}{dR}\frac{dR_s}{d\rho}
\left(\tensor{\rho}{_,_a}\tensor{dx}{^a
}\right)-\g v(dz-vdt)\frac{R_s}{\rho} \nonumber
\\
{}&\qquad{}-\frac{\g v(z-vt)}{\rho}\left(\frac{dR_s}{d\rho}-
\frac{R_s}{\rho}\right)\left(\tensor{\rho}{_,_a}\tensor{dx}{^a}\right) ,
\end{align}
and where the derivative of (\ref{eqn:rhosquared}) gives
\begin{equation}
\tensor{\rho}{_,_a}\tensor{dx}{^a}=
\frac{1}{\rho}\left[xdx+ydy+\g^2(z-vt)(dz-vdt)\right] .
\end{equation}
\end{subequations}

Finally, it is worth asking whether the sequence of steps in Table 
\ref{table:coordinatechanges} is essential.  For example, it might be 
possible to use the results of Hannam \emph{et al}.~\cite{HannETC08} as 
reviewed 
in \ref{subsec:HHOBO_results} to construct first an unboosted black hole 
with trumpet geometry and then apply a global Lorentz boost to the 
resulting metric in rectangular coordinates.  We do not presently know 
whether this alternative approach might work and it may well be worth 
future study.

\section{Setup for Simulations and Numerical Tests}
\label{sec:numerical_setup}

Our simulations were done using the Somerville release of the 
\texttt{Einstein Toolkit} (ET) \cite{ET}.  A new boosted trumpet initial data 
thorn was created based 
on the procedure outlined in the previous section.  In addition, we made use 
of a number of preexisting thorns, including (a modification of) 
\texttt{McLachlan} 
\cite{McLa} for evolution and \texttt{TwoPunctures} \cite{AnsoBrugTich11} to 
generate Bowen-York initial data for comparison simulations as controls 
(discussed below).  Several diagnostic thorns were utilized, including 
\texttt{AHFinderDirect} \cite{Thor96,Thor03} to identify apparent horizons,
(a modification of) \texttt{QuasiLocalMeasures} \cite{DreyETC03} to measure 
ADM mass and momentum and apparent horizon properties, \texttt{WeylScal4} 
\cite{BakeCampLous02} to compute the Weyl curvature quantities, and 
\texttt{Multipole} \cite{Mult} to generate spin-weighted spherical harmonic 
amplitudes of gravitational waveforms.  Various results using these thorns are 
shown in Sec.~\ref{sec:results}.  Post-processing of data was carried out with 
the help of the \texttt{SimulationTools} package for Mathematica \cite{SimT}.

To construct the initial data for a single boosted black hole, we start with 
Mathematica expressions for the components of the metric $\tensor{g}{_a_b}$ 
that produce the line element seen in (\ref{eqn:full_line_element}).  We then 
calculate symbolically the first and second derivatives of the metric, 
$\tensor{g}{_a_b_,_c}$ and $\tensor{g}{_a_b_,_c_d}$.  The expressions for the 
components of the metric and its derivatives are then written to a header 
file by Mathematica.  The lookup table for $dh_s/dR$ and $R_s$ (and their 
derivatives) versus $\rho$ is also computed and exported from Mathematica
(as described at the end of \S\ref{subsubsec:integral_ODE}).

Once the simulation mesh with its refinement levels has been constructed 
by \texttt{Carpet}, our initial data thorn (written in C) sweeps over 
the mesh using analytic expressions and interpolations of the numerical 
solutions for $dh_s/dR$ and $R_s$ to populate the arrays of required 3+1 
quantities.  Given a starting time $t$ (typically zero), at every spatial 
location the code computes $\rho$ based on the specified value of $v$ (and 
$\gamma$).  An interpolation of the lookup table is then made to find 
$R_s(\rho)$ and its first three derivatives and to find $dh_s(R_s(\rho))/dR$ 
and its first two derivatives.  The interpolation is done with a cubic spline 
from the Gnu Scientific Library \cite{GSL}.  (The higher derivatives of 
the numerical functions and
the second derivative of the metric are computed in order to assess constraint 
violations, especially in applications where we superpose two boosted trumpets 
to initiate a binary encounter.)

At each point interpolated numerical values are inserted in the 
Mathematica-exported expressions for $\tensor{g}{_a_b}$, 
$\tensor{g}{_a_b_,_c}$, and $\tensor{g}{_a_b_,_c_d}$, and from these we 
compute the determinants of the full metric ($g$) and spatial metric 
($\gamma$), the inverse spatial metric $\tensor{\gamma}{^i^j}$, and then the
remaining 3+1 quantities 
\begin{subequations}
\label{eqn:threeplusoneeqns}
\begin{align}
\alpha&=\sqrt{-g/\gamma} , \\
\tensor{\beta}{^i}&=\tensor{g}{_0_j}\tensor{\gamma}{^i^j} , \\
\tensor{\beta}{^i_,_t}&=\tensor{g}{_0_j_,_t}\tensor{\gamma}{^i^j}+
\tensor{g}{_0_j}\tensor{\gamma}{^i^j_,_t} , \\
\tensor{K}{_i_j}&=\frac{1}{2\alpha}\big[-\tensor{g}{_i_j_,_0}+
\tensor{g}{_0_j_,_i}+\tensor{g}{_0_i_,_j}\big. \nonumber \\
{}&\big.\qquad\qquad\qquad{}-\tensor{\beta}{^l}\left(\tensor{g}{_l_j_,_i}+
\tensor{g}{_i_l_,_j}-\tensor{g}{_i_j_,_l}\right)\big] .
\end{align}
\end{subequations}

\subsection{Grid Set Up}

For a typical boosted black hole simulation, we use \texttt{Carpet} to set 
up five levels of mesh refinement surrounding the black hole puncture beyond 
the coarsest adaptive mesh refinement (AMR) level (which covers the entire 
simulation domain).  A baseline grid would have $\Delta x=0.8M$ and 
$\Delta t=0.36M$ on the coarsest level, with the mesh spacing halved on each 
successive refinement level.  This gives $\Delta x = 0.025M$ on the finest 
level after five refinement steps.  On each level the time step is halved also 
so that the Courant-Friedrichs-Lewy parameter \cite {CourFrieLewy28} is 
maintained.  The half-lengths of the AMR cubes 
are $\{0.9,1.8,3.6,7.2,14.4\}M$ (not including ghost zones).  For tests of 
the effects of resolution, we scale the baseline grid $\Delta x$ and 
$\Delta t$ down by factors of $2/3$ and $4/9$ and up by a factor $3/2$.

For a single boosted black hole, we also utilize symmetry across $x=0$ and 
$y=0$ to reduce the computational task.  The simulation domain in those cases 
is $[0,112M]$ along the $x$- and $y$-axes and $[-112M,224M]$ along the 
$z$-axis.  When computing ADM quantities and spin-weighted spherical harmonic 
projections of Weyl scalars, we integrate over spherical surfaces (as many as 
four to ten of them) that are centered on the black hole's original location.  
These diagnostic surfaces have coordinate radii between $r = 33M$ and 
$r = 100M$, spaced evenly in $1/r$.

\subsection{Simulation Coordinate Conditions}
\label{subsec:simulation_gauge_conditions}

The numerical evolution uses the 1+log slicing condition
\begin{equation}
\label{eqn:onepluslog}
\left(\tensor{\partial}{_t}-\tensor{\b}{^i}\tensor{\partial}
{_i}\right)\a=-2 \a K ,
\end{equation}
(Eq.~\eqref{eqn:onepluslogn} with $n = 2$), which is consistent with the 
assumption made in our initial data construction.  For a static black hole, 
the advection term in \eqref{eqn:onepluslog} plays a key role in determining 
both the value of radius $R_0$ on the limiting surface of the trumpet and the 
steady state behavior of the trace of the extrinsic curvature $K$.  If the 
advection term were not included for example, then in steady state 
($\partial_t\alpha = 0$) the slices would become maximal ($K = 0$).  

\begin{figure}
\includegraphics[scale=1]{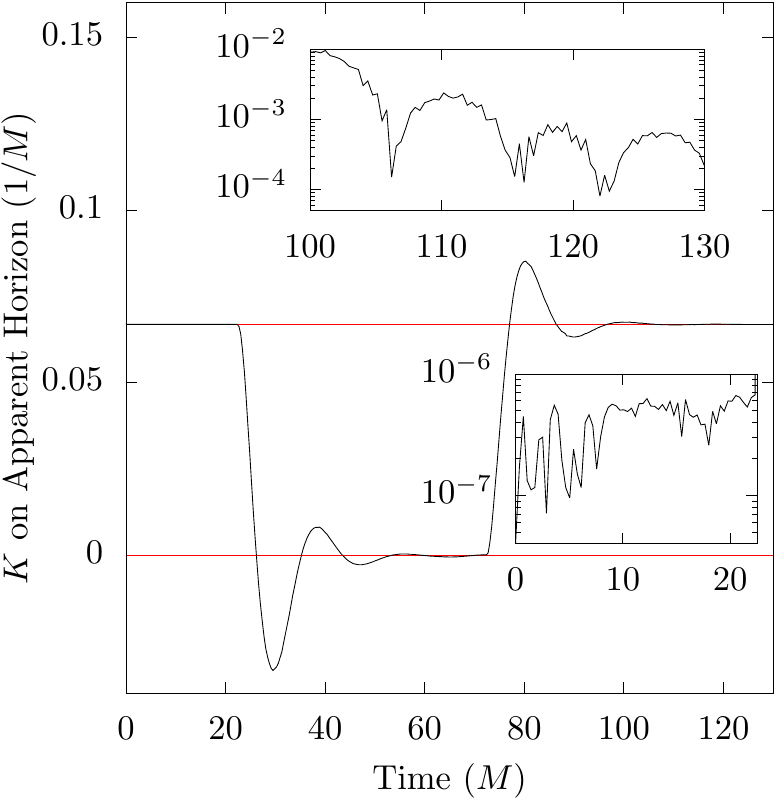}
\caption{Mean curvature $K$ at the apparent horizon of a static black 
hole versus time.  The simulation begins with advection in the 1+log slicing 
condition turned on.  Advection is turned off at $t=22.5M$ and back on again 
at $t=72.5M$.  During the switch-off $K$ is driven from the steady state 
value $K = 6.6856 \times 10^{-2}M^{-1}$ toward zero in a damped 
oscillation over 
several black hole light crossing times.  It accurately recovers (top inset) 
after advection is switched back on; insets show relative error. 
The results are consistent with 
\cite{HannETC08} (see their Figure~21) and used $\eta = 2/M$.
\label{fig:AHradius_changingadvection}} 
\end{figure}

An important first test of our initial data and configuration of the evolution 
code is to see that we recover the known behavior in the static case.  
Fig.~\ref{fig:AHradius_changingadvection} shows our accurate reproduction 
of a test made in Hannam \emph{et al}.~\cite{HannETC08} (their Figure 21) in which 
a static black hole is modeled and the advection term in 
\eqref{eqn:onepluslog} is turned off (at $t = 22.5 M$) and turned back on 
(at $t = 72.5 M$).  During the period when the advection term is switched 
off, $K$ at the apparent horizon is driven with a damped oscillation to 
zero.  The insets show the relative accuracy maintained in $K$ in the steady 
state periods before the term is switched off and after it is switched back 
on.  The advection term is utilized under normal circumstances.

\begin{figure}
\includegraphics[scale=1]{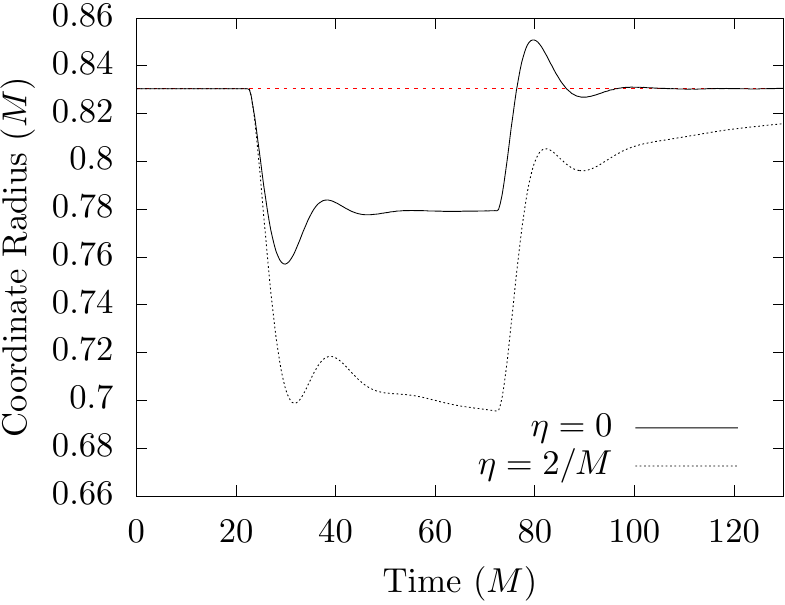}
\caption{Coordinate radius of the apparent horizon of a static black hole 
versus time.  The simulation begins with advection in the 1+log slicing 
condition turned on.  Advection is turned off at $t=22.5M$ and back on again 
at $t=72.5M$.  The apparent horizon suddenly shrinks in coordinate radius 
after $t = 22.5 M$, only to recover again after advection is restored.  
Two simulations are shown with $\eta = 0$ (solid) and $\eta = 2/M$ (dashed).  
\label{fig:AHradius_changingeta}} 
\end{figure}

The numerical evolution also uses the hyperbolic $\Gamma$-driver condition 
\cite{BaumShap10} for the shift vector
\begin{subequations}
\begin{align}
(\tensor{\partial}{_t}-\tensor{\b}{^j}
\tensor{\partial}{_j})\tensor{\b}{^i}&=\frac{3}{4}\tensor{B}{^i}\\
(\tensor{\partial}{_t}-\tensor{\b}{^j}\tensor{\partial}{_j})\tensor{B}{^i}&=
(\tensor{\partial}{_t}-\tensor{\b}{^j}\tensor{\partial}{_j})
\tensor{{\tilde{\G}}}{^i}-\eta\tensor{B}{^i} .
\end{align}
\end{subequations}
In typical numerical simulations, long term stability is improved by inclusion 
of the advection terms in this coordinate condition (making it the so-called 
``shifting shift'' condition \cite{BaumShap10}).  However, the simulation 
shown in Fig.~\ref{fig:AHradius_changingadvection} had the advection terms in 
the $\Gamma$-driver condition turned off in order to match the conditions used 
by Hannam \emph{et al}.~\cite{HannETC08}.  The parameter $\eta$ also has an 
effect 
on stability.  Typical values used are $\eta=0$, $\eta=1/M$, or $\eta=2/M$.
The previous test was also used to assess the effects of different choices of 
$\eta$.  Fig.~\ref{fig:AHradius_changingeta} shows the response in the 
coordinate radius of the apparent horizon in the test as the advection term 
in the 1+log slicing condition is turned off and then back on.  When advection 
is switched off the apparent horizon shrinks in coordinate radius, expelling 
mesh points from the interior.  Once advection is switched back on, the 
coordinate radius of the apparent horizon grows, seeking to recover its 
previous value.  Two different test simulations are shown, using $\eta = 2/M$ 
and $\eta = 0$, with results apparently matching those seen in Figure 22 of 
\cite{HannETC08}.  Coordinate adjustment is faster in the $\eta = 0$ case.  
The results shown in Fig.~\ref{fig:AHradius_changingadvection} for changes 
in $K$ came from the $\eta = 2/M$ test simulation.

\subsection{Bowen-York Initial Data as a Control}

In Sec.~\ref{sec:results} we examine various consequences of using the new 
boosted-trumpet initial data in numerical evolutions, making comparisons with 
the alternative of employing Bowen-York data.  For these comparisons 
Bowen-York initial data are generated by the \texttt{TwoPunctures} thorn 
\cite{AnsoBrugTich11}.  Constructing Bowen-York data for a single black hole 
requires some care however, since \texttt{TwoPunctures} is inherently designed 
to generate black hole binaries.  To obtain a single black hole we first set 
\texttt{TwoPunctures} to generate a black hole at the center of the mesh with 
an ADM puncture mass of $M = M_{+}^{ADM} = 1$ after having chosen the 
Bowen-York momentum to be $P_z^{+} = \gamma v M$ for a desired $v$.  Next, a 
second black hole with ADM puncture mass of $M_{-}^{ADM} = 10^{-4} M$ and 
momentum $P_z^{-} = 0$ is specified at a distance of $d=2\,000M$ from the 
origin.  Once \texttt{TwoPunctures} maps its solution to the actual 
computational domain, we end up with data for a single moving Bowen-York 
black hole superposed with a slight tidal field contribution that is below 
the level of truncation error ($M^\textrm{ADM}_-/d^3\simeq10^{-14} M^{-2}$).

\begin{figure}
\includegraphics[scale=1]{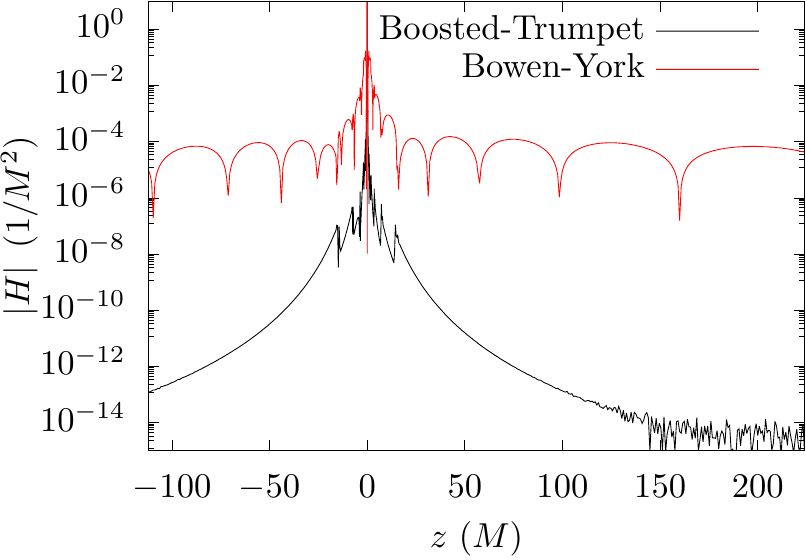}
\caption{Violations of the Hamiltonian constraint.  The violation is shown 
by plotting $\vert H \vert$ along the $z$-axis in both the boosted-trumpet 
(black) and Bowen-York (red) initial data.  In each case the black hole was 
given $v=0.5$ in the positive $z$ direction.  Large violations of the 
constraint near the puncture are confined within the horizon.}
\label{fig:H_violation}
\end{figure}

We can assess the accuracy in specifying initial data by either means by 
examining the degree of violation of the Hamiltonian constraint
\begin{equation}
H = R + K^2 - K_{ij} K^{ij} = 0 .
\end{equation}
In Fig.~\ref{fig:H_violation} we show the Hamiltonian constraint violation 
in the initial data by plotting $\log(\vert H \vert)$ along the 
$z$-axis.  Since the boosted-trumpet initial data for a single hole is derived 
from a partly-analytic, partly-numerical coordinate transformation of the 
Schwarzschild metric, it suffers very small Hamiltonian constraint violations 
over most of the mesh.  The larger Hamiltonian constraint violations in the 
Bowen-York data result from the level of convergence attained within the 
\texttt{TwoPunctures} routine.  As is well known, the constraint violation 
rises sharply near the puncture (see figure inset), due in part to difficulty 
in computing accurate finite differences of spatial curvature at that 
location.  These large deviations from the constraint are enclosed within 
the horizon.  The near-puncture violations of the Hamiltonian constraint are 
larger in the Bowen-York case, due again to (discretionary) limits set on 
convergence within \texttt{TwoPunctures}.  

\begin{figure}
\includegraphics[scale=1]{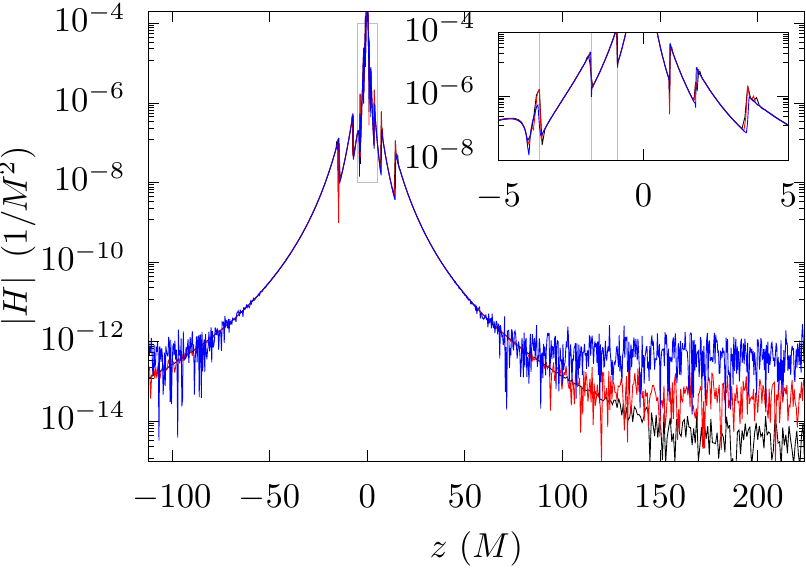}
\caption{Hamiltonian constraint violation scaled to show fourth order 
convergence. Plotted is $\vert H \vert$ along the 
$z$-axis for boosted-trumpet initial data at three resolutions.  The black 
curve has default grid spacing $\Delta x=0.8M$ on the coarsest level.  The 
red and blue curves have grid spacings scaled down by $\frac{2}{3}$ and 
$\frac{4}{9}$, respectively.  Fourth order convergence is demonstrated by 
multiplying the medium resolution curve by $(3/2)^4$ and 
the high resolution curve by $(9/4)^4$ so that they are coincident 
with the low resolution curve.  The black 
hole was set to have $v=0.5$ in the positive $z$ direction.  In the inset 
the locations of some AMR boundaries are indicated by vertical gray lines.
}
%\label{fig:H_rescaled}
\label{fig:H_resolution}
\end{figure}

The mesh resolution dependence of the Hamiltonian constraint violation in the 
boosted-trumpet initial data is shown in Fig.~\ref{fig:H_resolution}.  
In principle, a metric derived from a coordinate transformation of a 
Schwarzschild black hole should satisfy the constraints exactly.  The small 
levels of error seen here are primarily due to finite difference errors in 
computing the spatial curvature.  As such, the errors improve with increasing 
resolution.  This is demonstrated in the plot by scaling the errors under the 
assumption of fourth-order convergence.  The black curve corresponds to the 
standard grid spacing $\Delta x=0.8M$ on the coarsest level and the errors are 
not scaled.  The red and blue curves come from simulations which have grid 
spacings scaled down by 
$\frac{2}{3}$ and $\frac{4}{9}$, respectively; to demonstrate convergence 
their errors are scaled by the relevant mesh ratio to the inverse fourth 
power.  The boundaries of the AMR regions are visible, especially in 
the inset, caused by discontinuous changes in grid spacing.

\section{Results}
\label{sec:results}

\subsection{Qualitative Description of Boosted-Trumpet Data at $t=0$ and 
After Evolution}
\label{subsec:qualitative}

\begin{figure}[tbp]
\includegraphics[scale=1]{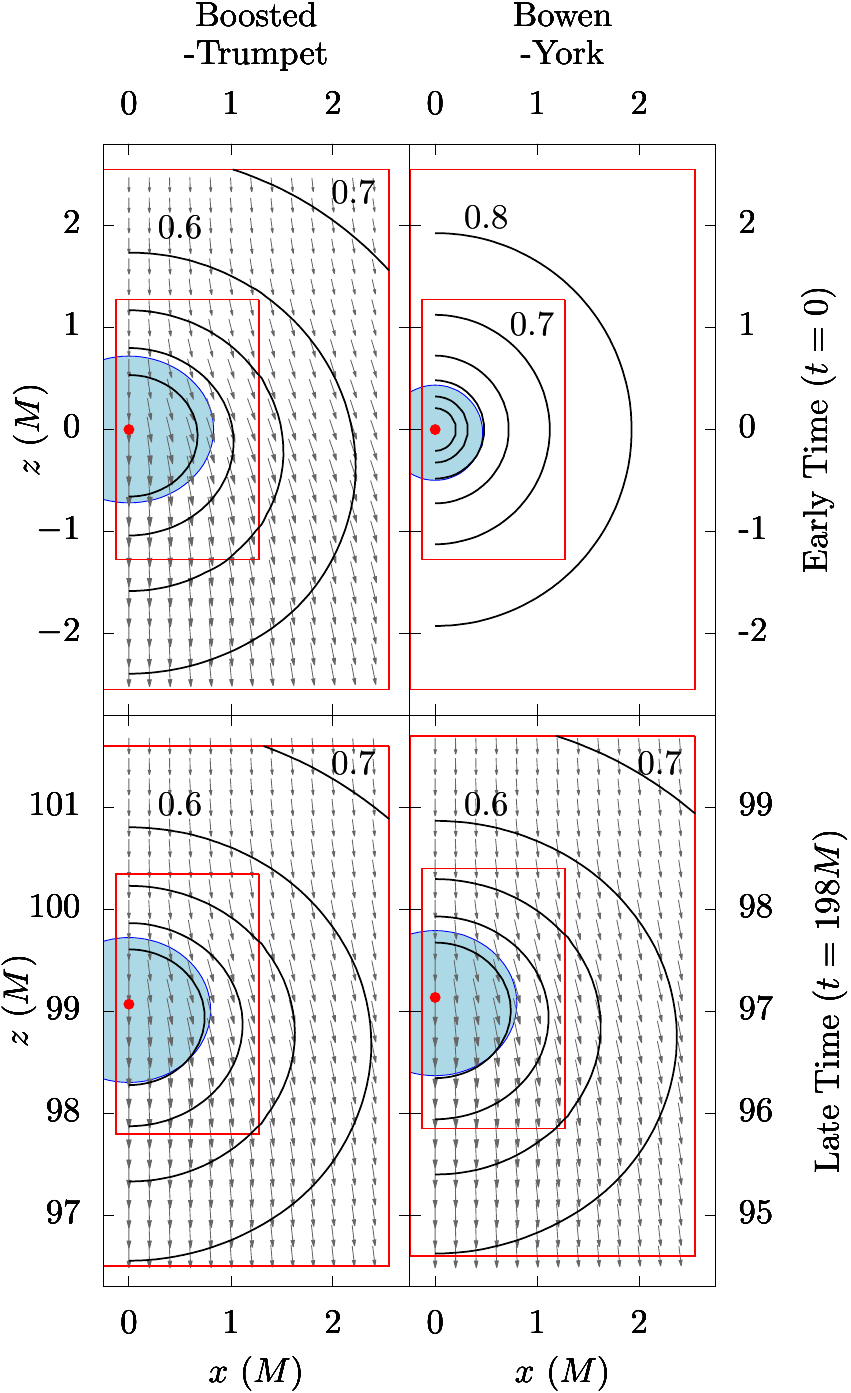}
\caption{Boosted-trumpet versus Bowen-York simulations.  Snapshots at $t=0$ 
(top panels) and at $t=198M$ (bottom panels) are shown.  The left panels 
depict the boosted-trumpet simulation while those on the right show the 
Bowen-York comparison.  Snapshots are a segment of the $x$,$z$ plane centered 
on the black hole and show (solid black) contours of the lapse, $x$ and $z$ 
components of the shift vector, trapped region (blue shaded) within the 
apparent horizon, and puncture location (red dot).  AMR boundaries are also 
shown (red box).  At $t=0$ the Bowen-York shift identically vanishes.  Lapse 
contours are spaced by $0.1$ with a couple of values labeled.  Each snapshot 
has equivalent spatial scale, though at late time are centered on different 
locations.  Each simulation had eventual steady-state black hole velocity of 
$v = 0.5$, but the Bowen-York hole had to accelerate from $v=0$ initially (see 
Fig.~\ref{fig:speed_from_shift}).  Simulations used $\eta = 0$.}
\label{fig:equatorial_slices}
\end{figure}

We begin the consideration of numerical evolution of the boosted-trumpet 
initial data with a plot of several 3+1 quantities in a planar cross section 
of the mesh.  A primary motivation for constructing initial data on a trumpet 
slice from the outset was the expectation that the initial data would be 
``closer'' to the steady state of the moving punctures gauge conditions than, 
for example, Bowen-York data.  Fig.~\ref{fig:equatorial_slices} shows this 
expectation to be borne out.  The plot makes a side by side comparison of a 
boosted-trumpet simulation (left side) and a Bowen-York simulation (right 
side), showing $t=0$ in the two top panels and $t=198M$ in the bottom two 
panels.  Each figure is a section of the $x$,$z$ plane, centered on the 
instantaneous position of the black hole.  Displayed within each figure are 
(solid black) contours of the lapse, arrows showing the $x$,$z$ components 
of the shift vector, the trapped region (blue shaded) within the apparent 
horizon, and a red dot showing the puncture location.  The (red) boxes show 
AMR boundaries.  In both cases the black hole parameters are set to give it 
an eventual steady-state velocity of $v = 0.5$ in the positive $z$-direction. 

At late times ($t\simeq 198M$) the metric near the black hole is very
similar in the two simulations, except the Bowen-York hole has not advanced 
as far in coordinate distance.  This is due in part to the Bowen-York hole 
having a vanishing shift vector $\tensor{\beta}{^i}$ initially 
(Fig.~\ref{fig:equatorial_slices}, upper right), and hence a vanishing initial 
puncture velocity.  The shift vector in 
the boosted-trumpet initial data is nonvanishing, and bears considerable 
resemblance to the steady-state shift at late time in the same simulation.  
The coordinate size of the apparent horizon in the Bowen-York initial data is 
smaller than its eventual steady-state dimensions.  There is much less change 
during the simulation in the coordinate size of the apparent horizon in the 
boosted-trumpet case.

\subsection{Quasi-local Energy and Momentum}
\label{subsec:quasilocal}

A primary reason for introducing trumpet slicing for initial data is to 
be able to better specify black hole momentum than is otherwise possible 
with Bowen-York data.  Momentum and energy of initial data can be determined 
in numerical relativity calculations using quasi-local measures of ADM momentum 
and ADM energy \cite{BaumShap10}.  Within the \texttt{Einstein Toolkit} there 
exists a routine \texttt{QuasiLocalMeasures} \cite{DreyETC03} for computing 
these estimates of the asymptotic quantities on two-surfaces of specified 
radius.  The ADM energy and momentum of Bowen-York black holes was studied 
previously \cite{YorkPira82,CookYork90}.  The conformally-flat nature of that 
data leads to the presence of spurious gravitational radiation, which adds to 
the ADM energy and affects the ability to specify the velocity (or Lorentz 
factor) of the black hole \cite{CookYork90,SperETC08}.

To make the analogous study of energy and momentum of boosted-trumpet initial 
data, we sought to make sharp estimates of (asymptotic) ADM energy and momentum 
by extrapolating the quasi-local measures.  To this end, we found it convenient 
to write somewhat simpler quasi-local energy and momentum integrals than are 
found in \texttt{QuasiLocalMeasures}.  We use
\begin{equation}
E_\textrm{ADM}(r)=
\frac{1}{16\pi}\oint_{r}\left[\tensor{\d}{^i^j}
\tensor{h}{_k_j_,_i}-\tensor{\partial}{_k}\left(\tensor{\d}{^i^j}
\tensor{h}{_i_j}\right)\right]\frac{\tensor{x}{^k}}{r}r^2d\O ,
\label{eqn:adm_mass_expression}
\end{equation}
and
\begin{align}
{}&\tensor{{P_\textrm{ADM}}}{^i}(r)=
\frac{1}{8\pi}\oint_{r}\tensor{\d}{^k^i}\bigg[
\frac{1}{2}\left(\tensor{h}{_k_0_,_j}+\tensor{h}{_j_0_,_k}-\tensor{h}{_j_k_,_t}
\right)\bigg.\nonumber\\
{}&\qquad\qquad\qquad\bigg.-\tensor{\d}{_k_j}\left(\tensor{h}{_l_0_,_l}-
\frac{1}{2}\tensor{h}{_l_l_,_t}\right)\bigg]\frac{\tensor{x}{^j}}{r}r^2d\O ,
\label{eqn:adm_momentum_expression}
\end{align}
where $\tensor{h}{_i_j}\equiv\tensor{g}{_i_j}-\tensor{\delta}{_i_j}$. 
In the limit $r\rightarrow\infty$ these expressions yield the ADM quantities: 
$E_\textrm{ADM}= E_\textrm{ADM}(\infty)$ and 
$\tensor{{P_\textrm{ADM}}}{^i}= \tensor{{P_\textrm{ADM}}}{^i}(\infty)$.  
These particular quasi-local formulae match those discussed in \cite{Alcu08} 
and we wrote a modified version of \texttt{QuasiLocalMeasures} for their 
use.  

By inserting the boosted-trumpet metric (\ref{eqn:full_line_element}) with the 
expansions from \S\ref{subsubsec:series_ODE} into 
(\ref{eqn:adm_mass_expression}) and (\ref{eqn:adm_momentum_expression}), and 
having Mathematica perform series expansions and integrals at $t=0$, we 
obtained asymptotic expansions for the ADM energy and momentum 
\begin{align}
E_\textrm{ADM}(r)&\simeq \g M + 
e_1(\gamma) \frac{M^2}{r} +
e_2(\gamma) \frac{M^3}{r^2} +
e_3(\gamma) \frac{M^4}{r^3} ,
\label{eqn:Eadm_expansion}
\\
P_\textrm{ADM}(r)&\simeq \g vM + 
p_1(\gamma) \frac{M^2}{r} +
p_2(\gamma) \frac{M^3}{r^2} +
p_3(\gamma) \frac{M^4}{r^3} ,
\label{eqn:Padm_expansion}
\end{align}
out to the indicated order.  Here the coefficients in the expansions are 
somewhat complicated expressions of $\gamma$ (or $v$), which for brevity we 
leave off listing.  It is apparent that asymptotically the initial data 
preserve the expected Christodoulou \cite{Chri70} energy-momentum relation 
$E_\textrm{ADM}^2 = P_\textrm{ADM}^2 + M^2$.  We also treated the 
coefficients as unknowns, measuring 
$E_\textrm{ADM}(r)$ and $P_\textrm{ADM}(r)$ at a minimum of four radii, and 
made fits to the expansions to cubic order in $1/r$.  Comparison between the 
numerically determined coefficients and their expected analytic values 
provided a strong check on the quasi-local measures, and the intercepts 
(at infinity) provided sharp, extrapolated estimates of energy and 
momentum.  Making this direct comparison is what necessitated introducing 
the modified versions of the ADM quantities 
(\ref{eqn:adm_mass_expression}) and (\ref{eqn:adm_momentum_expression}).

\begin{figure}
\includegraphics[scale=1]{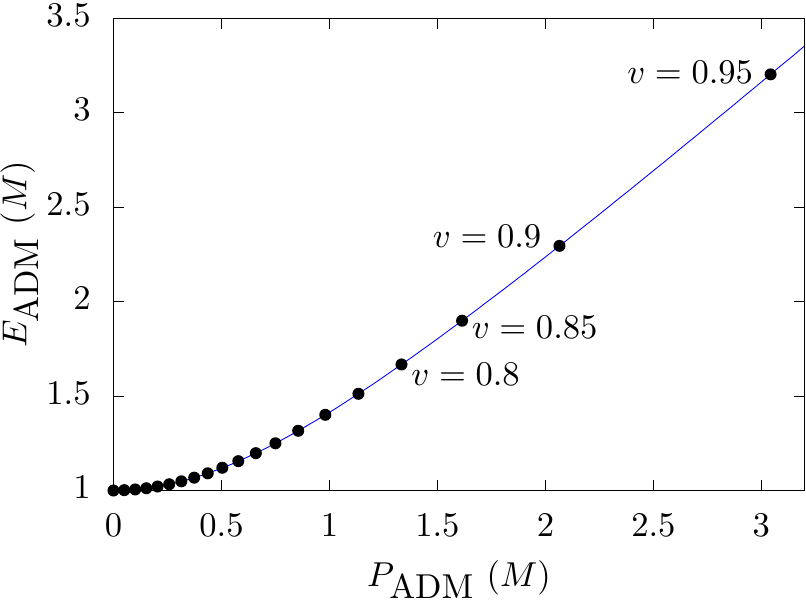}
\caption{ADM energy versus ADM momentum for boosted-trumpet initial data.  
Boosted-trumpet black hole initial data were computed for a set of 
velocities up to $v_\textrm{max}=0.95$.  Quasi-local estimates of ADM energy 
and momentum were computed on a set of two-surfaces with radii from $r=33M$ 
to $r=100M$.  Asymptotic expansions of energy and momentum were fit to 
determine estimates of ADM energy and momentum at spacelike infinity.  The 
resulting values are shown plotted against each other (black dots).  The blue 
curve displays the expected Christodoulou relationship
$E_\textrm{ADM}^2 = P_\textrm{ADM}^2 + M^2$ for a boosted black hole.}
\label{fig:mvspplot}
\end{figure}

The result of extracting the extrapolated estimates for boosted-trumpet data 
is shown in Fig.~\ref{fig:mvspplot}.  The extrapolated values of 
$E_\textrm{ADM}$ and $P_\textrm{ADM}$ are shown plotted against each other 
as a set of points for different prescribed boost parameters (with the last 
few points being marked by their associated velocity parameter).  These data 
are well fit (blue curve) by the expected hyperbolic relationship between 
$E_\textrm{ADM}$, $P_\textrm{ADM}$, and $M$.  Comparison can be made 
to the Bowen-York case \cite{CookYork90} (Figure 1 in that paper) where 
increasing (junk) gravitational radiation limited the range of specifiable 
black hole velocity.

\subsection{Reduced Junk Gravitational Radiation}
\label{subsec:psi_numerical}

Another primary motivation in developing boosted-trumpet initial data is to 
reduce or eliminate junk radiation.  Curvature in vacuum is characterized by 
the Newman-Penrose (or Weyl) scalars $\psi_0,\ldots,\psi_4$, defined by 
contracting the Weyl tensor on a suitable null tetrad.  When a 
quasi-Kinnersley tetrad $(l,k,m,\overline{m})$ is selected, the scalars 
$\psi_0$ and $\psi_4$ primarily measure ingoing and outgoing gravitational 
radiation, respectively, while $\psi_2$ represents the ``Coulombic'' part of 
the field.  To extract gravitational waves, we focus on $\psi_4$, which is 
defined by
\begin{equation}
\psi_4 = \tensor{C}{_a_b_c_d}\tensor{k}{^a}
\tensor{{\overline{m}}}{^b}\tensor{k}{^c}\tensor{{\overline{m}}}{^d},
\label{eqn:psi4_defn}
\end{equation}
where $\tensor{k}{^a}$ is the radially ingoing null vector and 
$\tensor{{\overline{m}}}{^a}$ is a complex angular null vector.  Then at 
sufficient distance from the source, the two metric perturbation polarizations 
(waveforms) are determined by the real and imaginary parts of $\psi_4$:
\begin{equation}
\psi_4 = \ddot{h}_{+} - i \ddot{h}_{\times} .
\end{equation}
Drawing upon the \texttt{Einstein Toolkit} we use the routines 
\texttt{WeylScal4} and \texttt{Multipole} to compute $\psi_4$ and extract 
from it its $s=-2$ spin-weighted spherical harmonic amplitudes as functions 
of time at fixed radii.  In \texttt{WeylScal4} the quasi-Kinnersley tetrad is 
obtained \cite{BakeCampLous02} using the full (simulation) metric, with the 
radial and angular directions defined relative to the center of the mesh.

\begin{figure}
\includegraphics[scale=1]{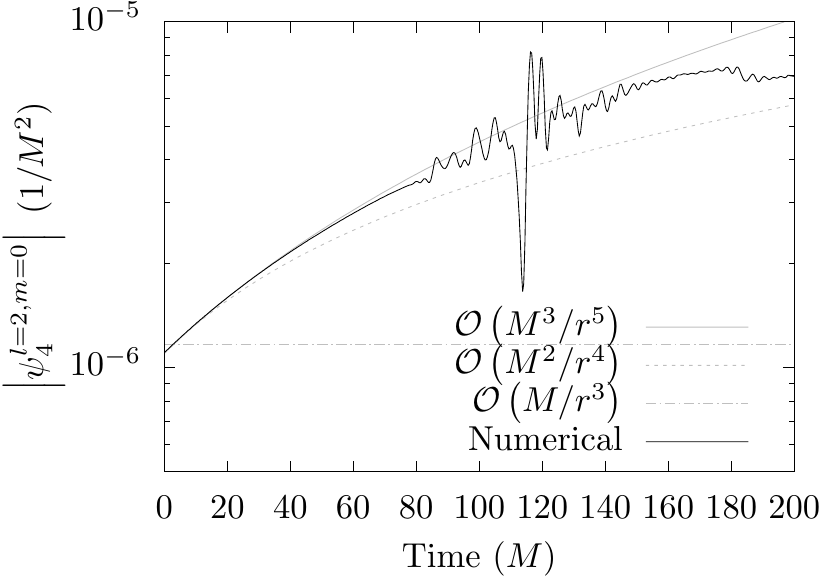}
\caption{Gravitational waveform from single boosted-trumpet black hole and 
underlying background.  Shown is the $l=2$, $m=0$ amplitude (black curve) of 
$\psi_4$ evaluated at a coordinate radius of $100 M$.  The junk waveform 
lies primarily in the interval $t = 80 - 140 M$.  Light dotted, dashed, and 
solid curves show successive approximations [given by 
(\ref{eqn:psi4_expansion})] to the underlying, nonradiative background in 
this signal (see text).  Simulation used $\eta = 1/M$.}
\label{fig:psi4_analytic_vs_numerical_r100_eta1}
\end{figure}

We use these tools to measure the junk gravitational radiation content of
our boosted-trumpet initial data.  Given axisymmetry, the $l=2$, $m=0$ mode 
dominates the radiation.  Fig.~\ref{fig:psi4_analytic_vs_numerical_r100_eta1} 
displays this amplitude (black curve) extracted at $r = 100 M$ as a function 
of $t$ from a simulation of a single black hole specified to have $v = 0.5$.
The time series contains two components, with the high frequency signal in 
the interval $t = 80 - 140 M$ being the junk radiation.  The high frequency 
waveform peaks at $t \simeq 110 M$, with the time delay indicating that the 
junk emerges from the region very near the black hole.  The size of this pulse 
of junk radiation is sufficiently small (to be established shortly) that the 
curve is dominated by a second component--a smooth underlying background whose 
source we discuss next. 

\subsubsection{Effects of Offset Tetrad}

In the rest frame of a Schwarzschild black hole and in the naturally 
associated Kinnersley tetrad $(l',k',m',\overline{m}')$, 
all of the Weyl scalars vanish except $\psi^{\prime}_2 = -M/R^3$.  Any 
radiation added as a perturbation would appear in $\psi^{\prime}_4$ and 
$\psi^{\prime}_0$.  When a boosted black hole is modeled two things affect 
this clean separation.  First, the tetrad constructed by \texttt{WeylScal4} 
$(l,k,m,\overline{m})$ will be in a frame boosted with respect to the 
static-frame tetrad.  Even at $t=0$ the frame components differ (e.g., 
\eqref{eqn:boostednull}) and the new frame components are linear combinations 
of those in the static frame.  Second, a single black hole does not remain 
centered on the mesh origin, so the linear transformation between frames is 
time dependent and as time proceeds the tetrad generated by the code rapidly 
ceases to reflect the principal null directions of the black hole.  The net 
effect is that the $\psi_4$ measured by the code will itself be a linear 
combination of Weyl scalars in the static frame
\begin{equation}
\psi_4 = {A_{(4)}}^{(4)} \psi^{\prime}_4 
+ {A_{(4)}}^{(2)} \psi^{\prime}_2 + {A_{(4)}}^{(0)} \psi^{\prime}_0 ,
\end{equation}
and will reflect a time-dependent mixing with the longitudinal field. 

To confirm that the underlying trend in 
Fig.~\ref{fig:psi4_analytic_vs_numerical_r100_eta1} is due to frame 
mis-centering and not radiation, we can take the initial data metric 
\eqref{eqn:full_line_element} for a boosted black hole, with its full time 
dependence, construct the curvature tensor, and then project the curvature on 
a tetrad constructed exactly as done in \texttt{WeylScal4}.  The initial data 
metric is exactly a Schwarzschild black hole, just in boosted trumpet 
coordinates, so it would have vanishing Weyl scalars in the Kinnersley 
frame except $\psi^{\prime}_2$ (i.e., no radiation).  In the mesh-centered 
frame, however, we find $\psi_4 \ne 0$. % This is not an exact analysis of 
%the effects seen in Fig.~\ref{fig:psi4_analytic_vs_numerical_r100_eta1} 
%because the metric in the simulation for $t>0$ follows the moving punctures 
%gauge conditions.
To compute this comparison we made an asymptotic 
expansion of the metric (\ref{eqn:full_line_element}), using in part the 
expansions for $h_s(R)$ and $R_s(\rho)$ from \S\ref{subsubsec:series_ODE}, to 
obtain expansions of $\psi_4$.  That expansion was in turn projected on the
$s=-2$ spin-weighted spherical harmonics, allowing us to extract the $l=2,m=0$ 
amplitude as an expansion in $1/r$ and as a function of time.  The 
calculation was done in Mathematica and yielded a complicated expression in 
terms of $v, M, r, t$.  Setting $v=0.5$ for the case considered in 
Fig.~\ref{fig:psi4_analytic_vs_numerical_r100_eta1}, we find 
\begin{align}
{}&\psi_4^{(2,0)}\left(v=\tfrac{1}{2} \right)=-\sqrt{\frac{5\pi}{2}}
\frac{M}{4r^3}\left(27\ln3-28\right)\nonumber\\
{}&\quad{}-\sqrt{\frac{5\pi}{2}}\frac{M}{
24r^4 } 
\Big[\left(225\sqrt{3}-142\pi\right)M+12\left(27\ln3-28\right)t\Big]
\nonumber\\
{}&\quad{}-\sqrt{\frac{5\pi}{2}}\frac{M}{1512r^5}
\Big[4\left(4493-960\sqrt{3}
\right)M^2\Big.\nonumber\\
{}&\qquad\Big.{}+126\left(130\pi-261\sqrt { 3 } \right)Mt+
378\left(27\ln3-28\right)t^2\Big]\nonumber\\
{}&\quad{}+\mathcal{O}(r^{-6}) .
\label{eqn:psi4_expansion}
\end{align}
It is clear that the claimed mixing occurs, even at $t=0$, and is made worse 
as the black hole shifts toward the extraction radius $r$.  

\begin{figure}
\includegraphics[scale=1]{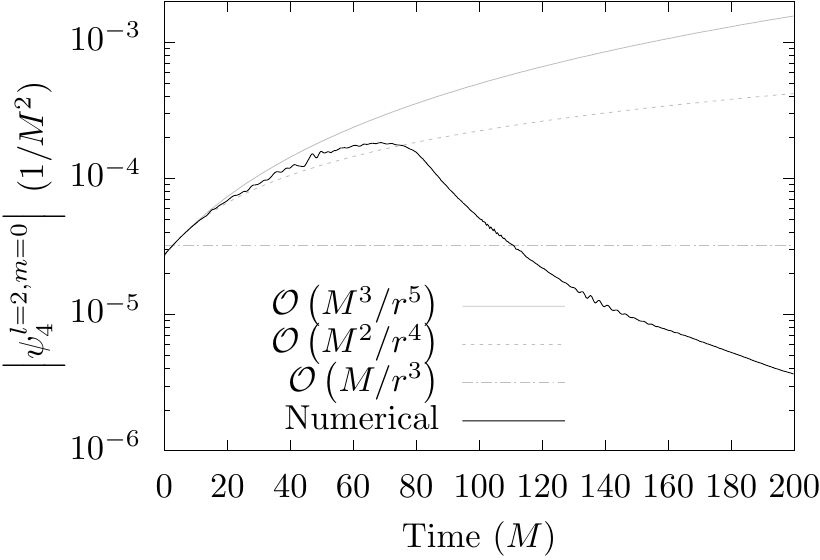}
\caption{Gravitational waveform from single boosted-trumpet black hole and 
underlying background.  Shown is the $l=2$, $m=0$ amplitude (black curve) of 
$\psi_4$ evaluated at a coordinate radius of $33M$.  The (small) junk 
waveform lies in the interval $t = 25 - 50 M$.  Light dotted, dashed, and 
solid curves show successive approximations [given by 
(\ref{eqn:psi4_expansion})] to the underlying, nonradiative background 
(see text).  The approximate explanation of this background breaks down 
beyond $t\simeq 66M$, as discussed in the text. Simulation used $\eta = 1/M$.}
\label{fig:psi4_analytic_vs_numerical_r33_eta1}
\end{figure}

In Fig.~\ref{fig:psi4_analytic_vs_numerical_r100_eta1} we show the progressive 
influence of terms in the expansion \eqref{eqn:psi4_expansion}, starting 
with the (constant) $M/r^3$ (dotted) term and adding in the $M^2/r^4$ 
(dashed) and $M^3/r^5$ (solid) terms.  Successive terms better approximate 
the smooth underlying trend.  However, our approximate analysis breaks 
down at late times, as can be seen in 
Fig.~\ref{fig:psi4_analytic_vs_numerical_r33_eta1}, once the black hole 
drifts into contact with and moves beyond the extraction radius and our 
asymptotic expansions are no longer valid.  The conclusion remains, though, 
that the junk radiation is the smaller high-frequency signal superimposed 
on the smooth non-radiative underlying trend.  The junk radiation is clearly 
seen in Fig.~\ref{fig:psi4_analytic_vs_numerical_r100_eta1}, where the 
extraction radius is $r=100 M$, but is barely noticeable in 
Fig.~\ref{fig:psi4_analytic_vs_numerical_r33_eta1}, where $r=33 M$ and 
the $r^{-3}$ longitudinal term dominates the $r^{-1}$ radiative term. 

\subsubsection{Boosted-Trumpet versus Bowen-York Junk Radiation}

Our method of constructing boosted-trumpet initial data greatly minimizes 
junk radiation, and its resulting small amplitude is the reason the radiation 
is dominated by the offset-tetrad effects in the preceding two plots.  To 
quantify this claim we ran a comparison simulation with Bowen-York initial 
data and with exactly the same prescribed black hole momentum.  The $l=2,m=0$ 
amplitude of $\psi_4$ was extracted similarly at $r=100 M$ and $r=33 M$ and 
overlaid on the comparison boosted-trumpet waveform at the same radii.

\begin{figure}
\includegraphics[scale=1]{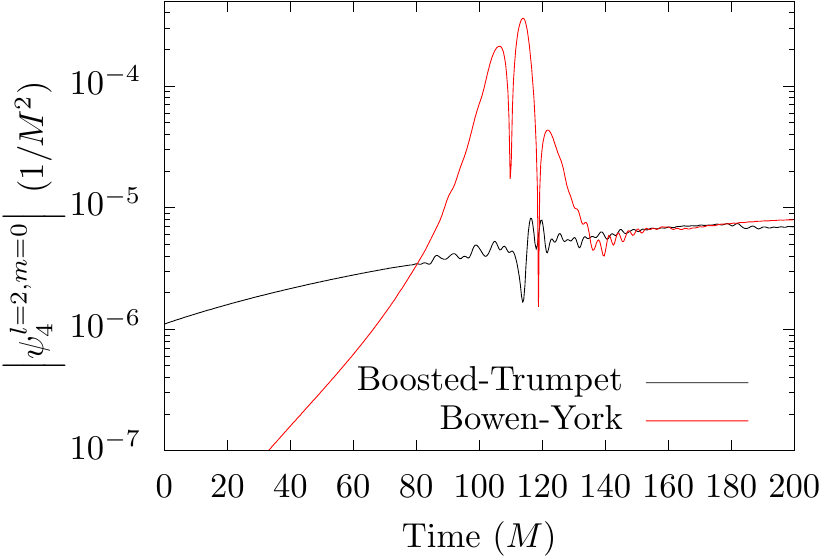}
\caption{Junk radiation comparison between simulations with boosted-trumpet 
and Bowen-York initial data.  The dominant $l=2,m=0$ amplitude of the Weyl 
scalar $\psi_4$ is extracted at $r=100 M$ in a boosted-trumpet simulation 
(black curve) and Bowen-York comparison (red curve).  Following passage of 
the pulse the waveforms match, reflecting the nonradiative offset-tetrad 
background.  Simulations used $\eta=1/M$. }
\label{fig:psi_r100}
\end{figure}

\begin{figure}
\includegraphics[scale=1]{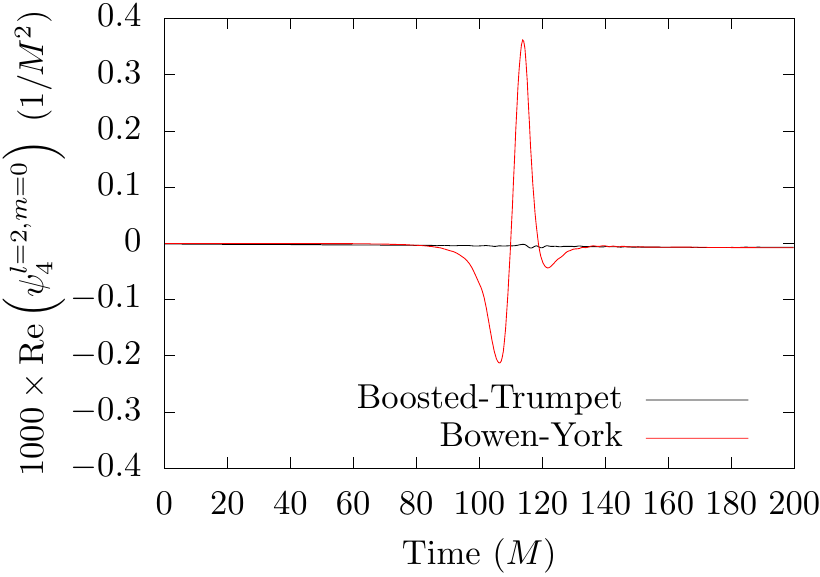}
\caption{Junk radiation comparison between simulations with boosted-trumpet 
and Bowen-York initial data.  Same as Fig.~\ref{fig:psi_r100} except on a 
linear scale.  Bowen-York signal (red curve) dominates over the barely 
visible wave (black curve) from the boosted-trumpet model.}
\label{fig:psi_r100_linear}
\end{figure}

Fig.~\ref{fig:psi_r100} shows the $l=2$, $m=0$ amplitude of $\psi_4$ as a 
function of time extracted at $r=100 M$ for both the $v=0.5$ boosted-trumpet 
black hole simulation (black curve) and the comparable Bowen-York model 
(red curve).  The Bowen-York junk radiation pulse centered about 
$t \simeq 110 M$ has an amplitude nearly two orders of magnitude larger than 
the boosted-trumpet pulse.  After the passage of the pulse, the waveform 
settles in both simulations to the nonradiative offset-tetrad background.  At 
early times, before the arrival of the pulse, the asymptotic behavior of the 
Bowen-York data differs from boosted-trumpet dependence, rendering our 
offset-tetrad analysis inapplicable until passage of the junk pulse.  
Fig.~\ref{fig:psi_r100_linear} shows the same comparison but on a linear 
scale, making even more evident the drastic reduction in junk radiation 
amplitude.

\begin{figure}
\includegraphics[scale=1]{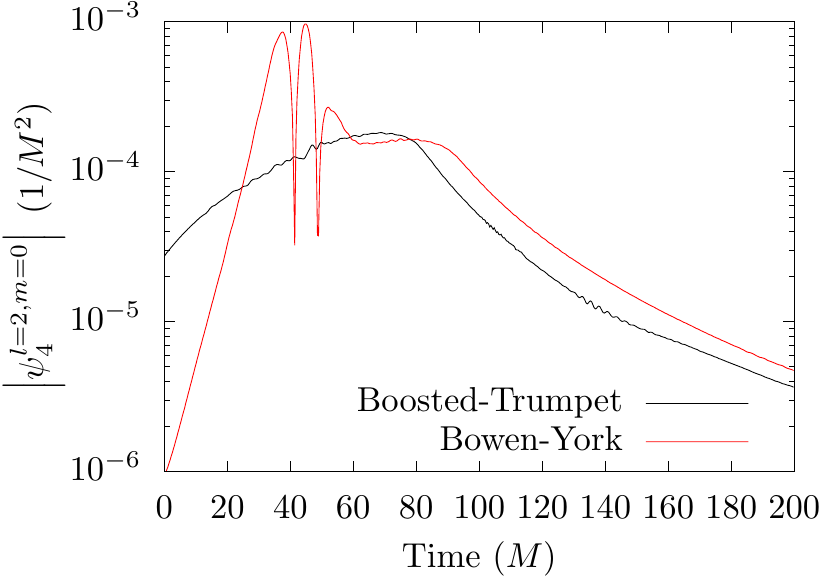}
\caption{Junk radiation comparison between simulations with boosted-trumpet 
and Bowen-York initial data.  Same as Fig.~\ref{fig:psi_r100} except the 
waveforms are extracted at a radius of $33 M$.  At late times following 
passage of the pulse, the nonradiative offset-tetrad background prevails, 
with a relative delay evident in the arrival of the Bowen-York black hole at 
the extraction radius due to early coordinate velocity changes.  Simulations 
used $\eta = 1/M$.}
\label{fig:psi_r33}
\end{figure}

Fig.~\ref{fig:psi_r33} shows the same comparison between $l=2,m=0$ amplitudes 
of $\psi_4$ but extracted at $r=33 M$.  It is still clear that the Bowen-York 
junk pulse is several orders of magnitude larger than the boosted-trumpet 
pulse, though the offset-tetrad effects are more pronounced at this radius.
At late times after the passage of the junk pulse, the two waveforms behave 
similarly.  The relative delay in the Bowen-York signal can be ascribed to 
a coordinate effect, as the Bowen-York black hole is delayed in reaching the 
extraction radius by about $10 M$ (in the $\eta = 1/M$ simulation) because it 
must accelerate from vanishing initial velocity.  

Finally, there is the matter of what sets the scale of junk radiation in 
the boosted-trumpet case and why it is present at all.  In principle our 
prescription for a single boosted-trumpet black hole is merely a coordinate 
transformation of the exact Schwarzschild spacetime, which contains no 
radiation whatsoever, and the adjustment to moving-punctures gauge is a 
coordinate effect.  This certainly explains why the boosted-trumpet junk 
is orders of magnitude smaller than that associated with Bowen-York data.  
The fact that the boosted-trumpet junk radiation is not exactly zero can be 
ascribed to numerical errors in constructing the initial data.  For example, 
we make numerical approximations for $dh_s/dR$ and $R_s$, and these 
approximations enter into the initial metric \eqref{eqn:full_line_element} 
in several places.  Additionally, components of the metric 
\eqref{eqn:full_line_element} are sampled at the mesh points and 
then other quantities, including finite differences for derivatives, are 
computed via \eqref{eqn:threeplusoneeqns} to round out the initial data.  
These numerical steps leave in their wake discretization and finite-difference 
errors.

\begin{figure}
\includegraphics[scale=1]{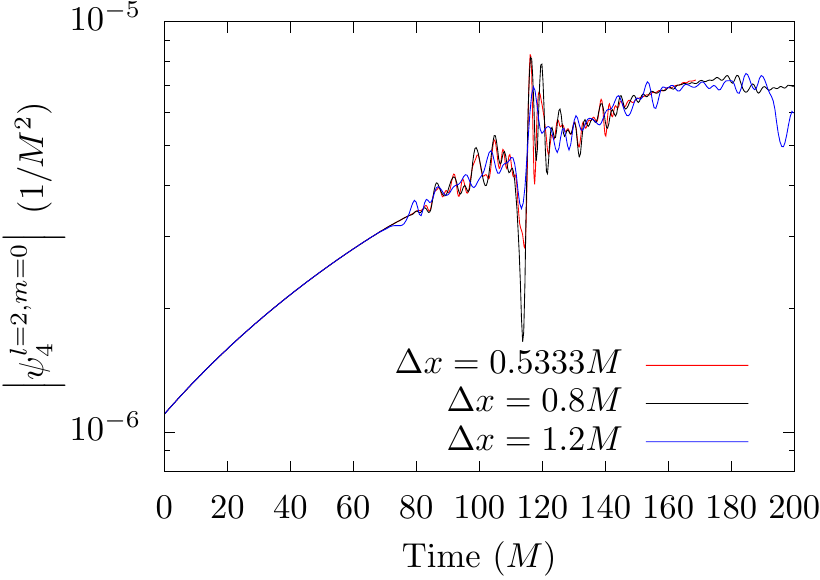}
\caption{Resolution dependence of the boosted-trumpet junk radiation and 
the offset-tetrad background waveform.  The same $l=2$, $m=0$ amplitude of 
$\psi_4$, extracted at $r=100 M$, is shown for three simulations with 
mesh resolutions indicated on the plot.  The early offset-tetrad effect is 
well-resolved, while the junk radiation signal is highly sensitive to changes 
in resolution, indicative of discretization and sampling errors.  
Simulations used $\eta = 1/M$.}
\label{fig:psi_resolution}
\end{figure}

What are the effects of changing mesh resolution?  
Fig.~\ref{fig:psi_resolution} shows the waveform extracted at $r=100 M$ for 
simulations with three different mesh resolutions.  The early part of the 
waveform due to the offset-tetrad effects is nearly insensitive to changes 
in resolution, indicating a nonvanishing and fairly well-resolved behavior.  
The higher frequency junk signal, on the other hand, is sensitive to changes 
in resolution without exhibiting a clear power-law scaling that would be 
expected of finite differencing a smooth function.  That lack of scaling 
likely stems from one or more sources.  First, as we scaled the computational 
mesh we did not also simultaneously scale the resolution on the spherical 
surfaces upon which the angular harmonic parts of $\psi_4$ are computed.  
Second, we also did not simultaneously scale the resolution of the lookup 
table.  Third, by construction, the junk radiation in our scheme should 
vanish analytically, which means that a numerical computation of $\psi_4$ 
will be subject to roundoff errors as projecting the Weyl tensor involves 
differencing nearly identical terms.  Finally, the small errors in the lookup 
table can show up in the boosted-trumpet metric in a stochastic fashion, 
since the tabulated spherical functions are sampled onto a rectangular grid.

\subsection{Coordinate Velocity of the Black Hole}

The Bowen-York prescription determines $g_{ij}$ and $K_{ij}$ within the 
initial spacelike slice but leaves unspecified the lapse $\alpha$ and shift 
$\beta_i$ both on and in the future of the initial surface.  Lacking any 
better choice the initial shift is frequently set to zero.  When used in a 
moving-punctures simulation, the mesh points near the Bowen-York center 
(other end of the wormhole) rapidly draw toward the trumpet surface and 
become points near the puncture (limit surface of the trumpet) 
\cite{HannETC08}.  One measure of the motion of the black hole is (minus) the 
value of the shift at the puncture.  As is well known, this has the odd, 
though purely coordinate, effect that in a Bowen-York simulation a black hole 
with initial linear momentum has no initial velocity.  We saw the effect in 
coordinate position/time delays both in Fig.~\ref{fig:equatorial_slices} and 
in Fig.~\ref{fig:psi_r33}.

In our boosted-trumpet procedure the full spacetime metric is constructed 
initially, which means that starting conditions for the lapse and (nonzero) 
shift are given.  Given the initial value of the shift at the puncture, the 
velocity of the black hole (as determined by the shift) will be specified 
correctly, consistent with the expected stationary value.  In practice, the 
puncture velocity varies in time somewhat, over a few light crossing times 
up to hundreds of $M$ depending upon $\eta$, before settling back to its 
original value.

\begin{figure}
\includegraphics[scale=1]{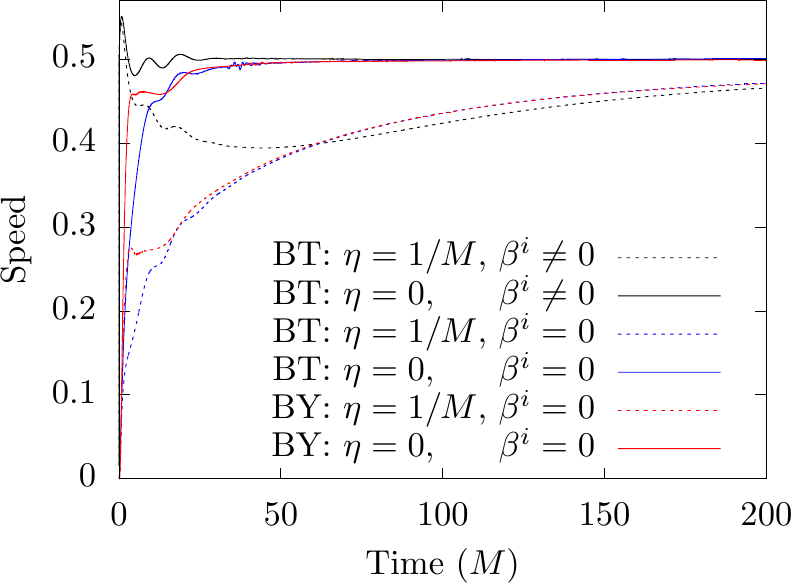}
\caption{Black hole speed computed from puncture value of shift versus time. 
Each simulation was initiated to have momentum consistent with $v=0.5$.  
Solid curves depict simulations with $\eta=0$.  BT denotes boosted-trumpet 
simulations, while BY corresponds to Bowen-York models.  The blue solid curve 
shows a boosted-trumpet model with $\beta_i = 0$ initially.  Dotted curves 
correspond to the same simulations but with the gauge condition set to 
$\eta=1/M$.}
\label{fig:speed_from_shift}
\end{figure}

Fig.~\ref{fig:speed_from_shift} shows black-hole coordinate speed as 
determined by the shift at the puncture for several boosted-trumpet and 
Bowen-York simulations, all with parameters set for steady state velocity 
$v=0.5$.  In the boosted-trumpet simulation (solid black curve) made with 
$\eta=0$, the black hole's speed is very stable and settles to 
its specified value as the coordinates relax over a few light crossing 
times.  In the comparable Bowen-York simulation (solid red) the black hole 
abruptly accelerates and reaches its stationary coordinate speed in just 
slightly longer time.  The velocity profile of the Bowen-York black hole is 
primarily due to the initially vanishing shift, as can be seen by a third 
simulation (solid blue) where we set the initial boosted-trumpet shift to 
zero.  The $\eta$ parameter in the $\Gamma$-driver condition plays a large 
role in the evolution of the black hole velocity.  The three simulations 
just described (with $\eta=0$) were repeated (dotted curves) after changing to 
$\eta = 1/M$.  Black hole coordinate velocity (as measured by the shift at 
the puncture) reaches its asymptotic value much more rapidly when $\eta=0$.  
This is consistent with the behavior in Fig.~\ref{fig:AHradius_changingeta} 
where the apparent horizon radius recovers from gauge changes more rapidly 
when $\eta=0$.

\begin{figure}
\includegraphics[scale=1]{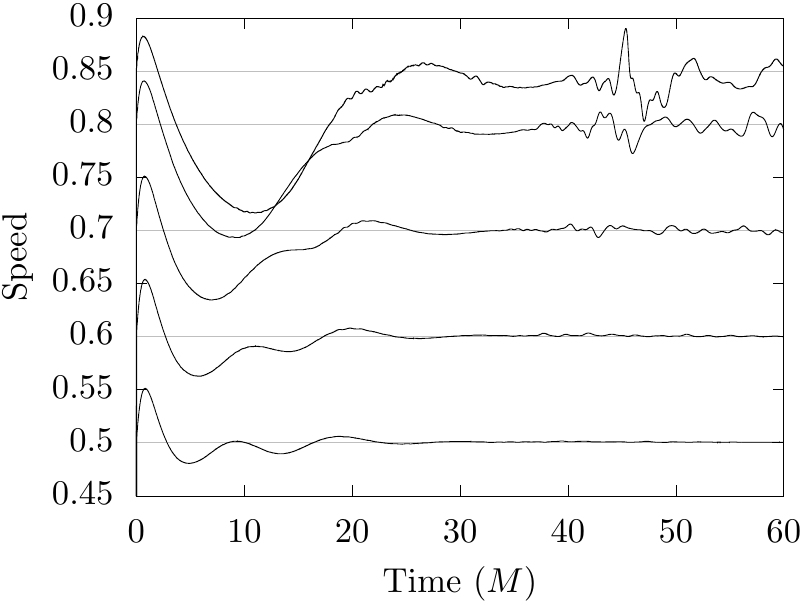}
\caption{Puncture velocity for boosted-trumpet black holes versus time for 
various specified speeds.  Specified initial and asymptotic speeds are 
indicated by the light horizontal lines.  Readjustment of the trumpet shape 
and settling of the spatial coordinates occur during $t\lesssim25 M$.  Simulations 
used $\eta = 0$. }
\label{fig:BT_speeds}
\end{figure}

Fig.~\ref{fig:BT_speeds} shows the black hole puncture-velocity history for 
a set of boosted-trumpet simulations with increasing Lorentz boosts.  Settling 
of the coordinates primarily occurs within the first $25 M$ in time as the 
black hole coordinate velocities tend toward their specified values (light 
horizontal lines).  The highest velocity model corresponds to Lorentz factor 
$\gamma = 1.90$.  Beyond this boost, we encountered problems with stability.  
The instability might result from the true steady-state trumpet being too 
distorted relative to our initially assumed spherically symmetric trumpet.  
Alternatively, we may merely need more mesh resolution at Lorentz factors 
$\gamma > 1.9$ than we were able to employ in this study.

\begin{figure}
\includegraphics[scale=1]{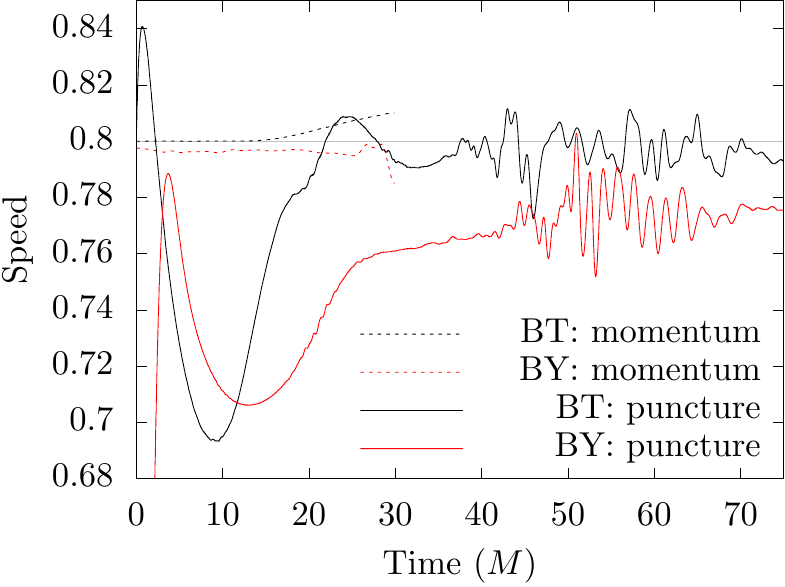}
\caption{Measures of black hole velocity at prescribed Lorentz factor 
$\g = 1.667$ ($v=0.8$).  Black curves correspond to the boosted-trumpet case and 
red curves to the Bowen-York.  Dotted curves estimate the velocity at early times 
using ADM momentum [see (\ref{eqn:lorentz_from_Padm})].  Solid curves show 
black hole velocity measured by the shift vector at the puncture.  
Simulations used $\eta=0$.  }
\label{fig:speed_from_shift_and_lorentz_080}
\end{figure}

A closer comparison between boosted-trumpet and Bowen-York black holes at 
high Lorentz factor can be seen in 
Fig.~\ref{fig:speed_from_shift_and_lorentz_080}.  The boosted-trumpet black 
hole (black curves) was specified by directly setting the boost velocity 
parameter to be $v=0.8$, or equivalently setting $\gamma = 1.667$.  The 
Bowen-York simulation was initialized by requiring the ADM mass at the 
puncture be $M=1$ and setting the Bowen-York momentum parameter to be 
$P = \gamma v = 1.333$.  Once the simulations are begun, we monitor the 
apparent horizon mass (a close proxy for the irreducible mass $M_{irr}$) and 
the (extrapolated to infinity) ADM momentum $P_{\rm ADM}$.  The latter two 
measures are then combined to yield an asymptotically-sensed estimate of the
Lorentz factor,
\begin{equation}
\g=\sqrt{1+\left(\frac{P_\textrm{ADM}}{M_\textrm{irr}}\right)^2}
\label{eqn:lorentz_from_Padm} ,
\end{equation}
and from that the velocity $v$.  This estimate of the velocity is plotted in 
Fig.~\ref{fig:speed_from_shift_and_lorentz_080} (dotted curves) at early times 
($t<30 M$) for both the boosted-trumpet (black) and Bowen-York (red) cases.  
(By $t=30 M$ and later the combination of outward-moving junk radiation and 
drift of the black hole toward the inner extraction radius makes extrapolated 
estimates of $P_{\rm ADM}$ unreliable.)  In the boosted-trumpet case, 
this measure of $v$ is indistinguishable at the 
resolution of the plot from the parameter choice in the initial data.  In the 
Bowen-York case, the asymptotic estimate of $v$ is a few thousandths smaller 
than the prescribed velocity, primarily because of the difference between the 
initial value parameter $M=1$ and the actual values of $M_{irr}$.

As an alternative, we plot in Fig.~\ref{fig:speed_from_shift_and_lorentz_080}
the puncture velocity of the black holes (solid curves) extracted from the 
two comparison simulations.  Once transient gauge effects have decayed the 
boosted-trumpet puncture velocity averages close to $v=0.8$ (solid black 
curve).  The late-time puncture velocity in the Bowen-York case averages 
about $v=0.78$ (solid red curve).  When converted to energy, this deficit 
corresponds to the Bowen-York black hole having about 4\% less energy than 
the boosted-trumpet hole, a drop ascribable to momentum and energy carried 
off by junk radiation and consistent with the undesired renormalization of 
initial momentum seen previously \cite{SperETC08} when using the Bowen-York 
prescription.  

\subsection{Apparent Horizon Comparisons}

We made further comparisons between boosted-trumpet and Bowen-York simulations 
by looking at circumferences on the apparent horizon.  Using 
\texttt{AHFinderDirect} first to locate the apparent horizon as a function 
of time and then using \texttt{QuasiLocalMeasures}, we computed the (maximum) 
proper circumferences of the apparent horizon in the $xy$-, $xz$-, and 
$yz$-planes
\begin{subequations}
\label{eqn:circumference_definitions}
\begin{align}
C_{xy}&=
\int^{2\pi}_0d\phi\sqrt{\tensor{q}{_\phi_\phi}\left(\th=\pi/2,\phi\right)}\\
C_{xz}&=2\int^\pi_0d\th\sqrt{\tensor{q}{_\th_\th}\left(\th,\phi=0\right)}\\
C_{yz}&=2\int^\pi_0d\th\sqrt{\tensor{q}{_\th_\th}\left(\th,\phi=\pi/2\right)} .
\end{align}
\end{subequations}
Here $\tensor{q}{_A_B}$ is the 2-metric on the apparent horizon
\begin{equation}
\tensor{q}{_A_B}=\frac{\partial\tensor{x}{^i}}{
\partial\tensor{\th}{^A}}\frac{\partial\tensor{x}{^j}}{
\partial\tensor{\th}{^B}}\tensor{g}{_i_j} ,
\label{eqn:2metric_def}
\end{equation}
where $A,B\in\{\th,\phi\}$.  Just like the apparent horizon, these proper 
lengths are not gauge invariant but depend upon the slicing condition.  For 
a Schwarzschild black hole in static coordinates these circumferences would 
all be equal, so any ratio would be unity.  

\begin{figure}
\includegraphics[scale=1]{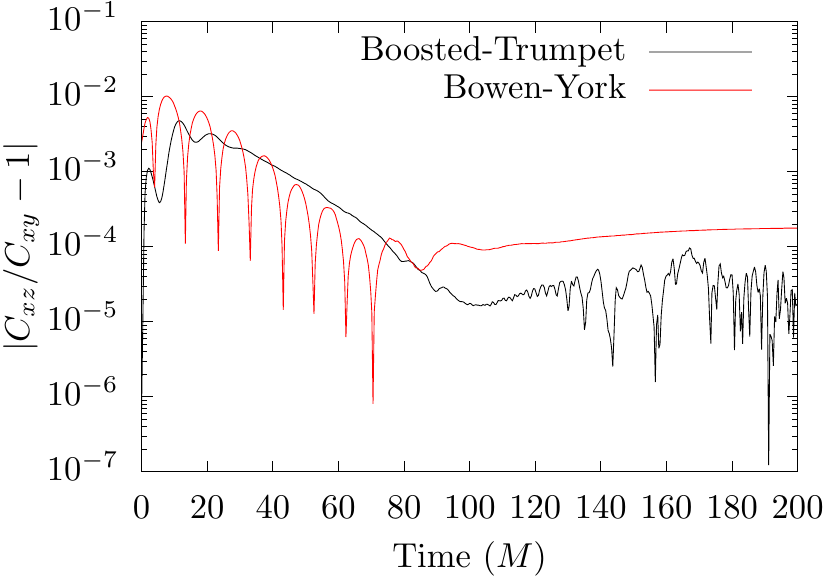}
\caption{Ratios of circumferences of the apparent horizon as functions of 
simulation time.  Both boosted-trumpet (black) and Bowen-York (red) black 
holes are set for $v=0.5$.  The damped oscillation in the Bowen-York case is 
suggestive of excitation of quasinormal ringing caused by the junk radiation.  
The boosted-trumpet distortion likely arises from evolution in the time 
slices as the trumpet shape and coordinates adjust.  Simulations used 
$\eta = 1/M$. }
\label{fig:AH_distortion}
\end{figure}

In Fig.~\ref{fig:AH_distortion} we plot the absolute value of the difference 
between the aspect ratio $C_{xz}/C_{xy}$ and unity on a log scale for 
simulations of $v=0.5$ black holes.  In the Bowen-York case (red curve) there 
is an initial distortion at about the 1\% level that subsequently decays by two 
orders of magnitude in a damped oscillatory motion.  This oblate to prolate 
oscillation suggests excitation of black hole quasinormal ringing, excited by 
the junk radiation, though with the ringing sensed locally in the apparent 
horizon properties and not remotely in the gravitational waveform.  In 
contrast, in the boosted-trumpet case (black curve) there is an initial 
distortion followed by rapid decay, but one lacking in repeated oscillations 
about zero.  In this case the significantly decreased presence of junk 
radiation bars noticeable quasinormal excitation of the apparent horizon.  A 
distortion is still present, because of the inconsistency between our initial 
trumpet shape and the steady state trumpet that emerges at late times in 
moving-punctures gauge, which decays as the time slices adjust.

\section{Conclusions}
\label{sec:conclusions}

We have demonstrated a means of constructing boosted-trumpet initial data 
for single black holes for use in moving-punctures simulations. 
%The method 
%can be extended to widely separated multiple black holes by superposing 
%initial data.
The procedure uses (1) a global Lorentz boost applied to Kerr-Schild 
coordinates, followed by (2) changing the time slice to trumpet topology by 
adding an appropriate height function, and (3) changing spatial coordinates 
to map the trumpet limit surface to a single (moving) point (i.e., the 
puncture).  This method can
be extended to widely separated multiple black holes by superposing
initial data; it could also form the basis for a prescription for specifying 
initial data for closer binaries as the initial guess for a re-solving of the
constraint equations.

We showed in simulations that the boosted-trumpet initial data 
more closely approximates eventual steady state geometry than does Bowen-York 
initial data.  Asymptotic ADM energy and momentum and apparent horizon mass 
follow closely the Christodoulou relationship, allowing large black hole 
velocities to be assigned (tested as high as $v=0.95$). 

Gravitational junk radiation is suppressed in simulations using the new 
scheme by two orders of magnitude relative to simulations with Bowen-York 
data of comparable parameters.  The essential element in reducing the junk 
radiation was use of non-conformally-flat boosted Kerr-Schild geometry
as an intermediate step in constructing the new data, with a trumpet 
introduced to ensure that the slice avoids the future singularity.  What junk 
radiation remains in the boosted-trumpet case is sensitive to changes in 
mesh resolution, reflecting 
an origin in sampling numerical data from a lookup table and resulting 
discretization errors.  Black holes initiated with the 
boosted-trumpet scheme have nonzero initial puncture velocities consistent 
with the intended boost.  As simulations begin, the puncture velocity varies 
for several black hole light crossing times before settling back to the 
intended value.  In the Bowen-York case, the eventual average puncture 
velocity is reduced slightly relative to the initial parameter value, 
reflective of some energy and momentum carried off in the junk radiation.
We also studied changes in the proper circumferences of the apparent horizon 
as functions of time.  The shape of the apparent horizon in the Bowen-York 
simulations suffers a damped oscillatory motion, suggestive of excitation 
of black hole quasinormal modes.  In contrast there is no noticeable ringing 
in the aspect ratio of the apparent horizon in boosted-trumpet simulations, 
though a slice-dependent initial distortion is seen to exponentially decay 
away.  

Given the reduction in both junk radiation and initial gauge dynamics, binary 
data based on this approach may facilitate more accurate simulations and 
gravitational waveforms with generically less gauge artefacts. Such improvements
in waveform accuracy will become increasingly important as gravitational-wave
detectors improve in sensitivity, for example with third-generation ground-based
detectors~\cite{Punturo:2010zz}, or the space-based LISA project~\cite{AmarETC17}.

\acknowledgments

We thank Ian Hinder, Seth Hopper, and Barry Wardell for helpful assistance 
with the \texttt{Einstein Toolkit} and with \texttt{SimulationTools}.  We 
also thank an anonymous referee for a detailed review that helped us 
clarify several important results.  This work was supported in part by NSF 
Grants No. PHY-1506182 and No. PHY-1806447.  K.S.~gratefully acknowledges 
support 
from the North Carolina Space Grant's Graduate Research Assistantship Program 
and a Dissertation Completion Fellowship from the UNC Graduate School.  
C.R.E.~acknowledges support from the Bahnson Fund at the University of North 
Carolina-Chapel Hill.  Simulations were run on XSEDE supercomputers using 
an allocation under Project No. PHY160024.

\bibliography{Boosted_Trumpet_Paper}

\end{document}